\documentclass[preprint,12pt]{elsarticle}

\usepackage{amssymb}
\usepackage{amsmath}
\usepackage{xcolor}
\usepackage{comment}
\usepackage{placeins}
\usepackage{hyperref}

\usepackage{color} 
\newcommand{\bl}{\color{blue}} 

\journal{Journal of Economic Dynamics and Control}

\begin{document}

\begin{frontmatter}



\title{A data-driven econo-financial stress-testing framework to estimate the effect of supply chain networks on financial systemic risk.}


\author[inst1,inst2]{Jan Fialkowski}
\author[inst1,inst5,inst6]{Christian Diem}
\author[inst1,inst3,inst5]{András Borsos}
\author[inst1,inst2,inst4]{Stefan Thurner\corref{Cor}}
\cortext[Cor]{Corresponding author at: Complexity Science Hub, Metternichgasse 8, A-1030, Austria. \ead{stefan.thurner@meduniwien.ac.at}}
\affiliation[inst1]{organization={Complexity Science Hub Vienna},
            addressline={Metternichgasse 8}, 
            postcode={A-1030}, 
            state={Vienna},
            country={Austria}}

\affiliation[inst2]{organization={Institute of the Science of Complex Systems, CeDAS, Medical University Vienna},
            addressline={Spitalgasse 23}, 
            postcode={A-1090}, 
            state={Vienna},
            country={Austria}}

\affiliation[inst3]{organization={Financial Systems Analysis, Central Bank of Hungary},
            addressline={Szabadság tér 9}, 
            city={Budapest},
            postcode={1054}, 
            country={Hungary}}

\affiliation[inst4]{organization={Santa Fe Institute},
            addressline={1399 Hyde Park Road},
            city={Santa Fe},
            postcode={NM 87501}, 
            country={USA}}

\affiliation[inst5]{organization={Institute for New Economic Thinking},
            addressline={Manor Road Building, Manor Road},
            city={Oxford},
            postcode={OX1 3UQ}, 
            country={United Kingdom}}

\affiliation[inst6]{organization={Smith School of Enterprise and the Environment},
            addressline={University of Oxford, South Parks Road},
            city={Oxford},
            postcode={OX1 3QY}, 
            country={United Kingdom}}

\begin{abstract}
Supply chain disruptions constitute an often underestimated risk for financial stability. As in financial networks, systemic risks in production networks arises when the local failure of one firm impacts the production of others and might trigger cascading disruptions that affect significant parts of the economy. Here, we study how systemic risk in production networks translates into financial systemic risk through a mechanism where supply chain contagion leads to correlated bank-firm loan defaults. We propose a financial stress-testing framework for micro- and macro-prudential applications that features a national firm level supply chain network in combination with interbank network layers. The model is calibrated by using a unique data set including about 1 million firm-level supply links, practically all bank-firm loans, and all interbank loans in a small European economy. As a showcase we implement a real COVID-19 shock scenario on the firm level. This model allows us to study how the disruption dynamics in the real economy can lead to interbank solvency contagion dynamics. We estimate to what extent this amplifies financial systemic risk. We discuss the relative importance of these contagion channels and find an increase of interbank contagion by 70\% when production network contagion is present. We then examine the financial systemic risk firms bring to banks and find an increase of up to 28\% in the presence of the interbank contagion channel. This framework is the first financial systemic risk model to take agent-level dynamics of the production network and shocks of the real economy into account which opens a path for directly, and event-driven understanding of the dynamical interaction between the real economy and financial systems.
\end{abstract}

\begin{highlights}
\item Systemic risk in a production network arises when a firms defaults causes other firms to stop production as well.
\item Systemic events on the production network can cause financial systemic risk through firm-credit defaults.
\item Interbank connections give rise to financial systemic risk and can further amplify supply chain contagion induced systemic risk.
\item We build a two-layer stress testing framework for micro- and macroprudential uses.
\item For the first time this allows us to study the interplay of interbank contagion and production network contagion
\end{highlights}

\begin{keyword}
financial stress-testing \sep 
supply chain networks \sep 
contagion channel interaction \sep systemic risk \sep financial networks \sep multi-layer networks
\JEL G210 \sep D200
\end{keyword}

\end{frontmatter}


\section{Introduction}
\label{sec:Intro}

The COVID-19 pandemic has highlighted the vulnerability of increasingly interconnected economic systems. The event caused extensive economic damage, especially for small and medium enterprises \citep{Bar20a}. To avoid more extreme damage and the spreading of it, government and central bank interventions were extensively implemented \citep{Mil24, ECB23}. The accompanying debates demonstrated the need for a better understanding of how these systems are connected. External events, such as climate change not only increases the risk of new epidemic diseases \citep{Bar19, Mar21, Rod21} but also the frequency and severity of natural disasters that have the potential to cause substantial economic damage \citep{Bat21, Wil18}. Large parts of these damages are mediated through supply chain or financial interconnections \citep{Hal08, Car21, Pic22}. Wars, trade wars, and changing political climate pose additional sources of risk for economic and financial systems.

By now it is well established that the interconnectivity of financial institutions is a significant source of systemic risk \citep{Bas11,Gla16,ECB23,Bar21,Bos04a,Ace15}. Different types of interconnectivity lead to different contagion channels through which systemic risk is created. Early research focused on simple contagion channels, 
such as asset-liability networks between financial institutions that can cause default cascades through solvency contagion \citep{Eis01,Bos04,Bos04a,All00,Gla16}. A proper implementation of default events lead to the so-called DebtRank \citep{Bat12, Thu13} that has been used to study the stability of financial systems \citep{Pol15, Bat16a, Car24,Wie23, Bar17}. Quadratic form variations of it were used to find financial networks with minimal systemic risk in financial systems \citep{Thu13,Pic21a,Die20}. Common asset holdings (overlapping portfolios) that cause contagion through fire sales \citep{Lev15,Con17,Fei17, Pic21a,Pol21} is another studied channel. 
Underlining the importance of understanding systemic risk and contagion channels from financial interlinkages, financial regulation has been taking this phenomenon into account for more than a decade\citep{Bas11,Ach14,Bor14,Arn12,Thu22}. 

Even though it is known that focusing on single contagion channels (e.g., interbank contagion or fire sales) in isolation can severely underestimate systemic risk \citep{Pol15, Det21, Wie23,Cac15}, most analyses of financial stability are performed on single contagion channels\citep{Eis01, Bat12, Die20, Gla16}. 
Financial systemic risk does not arise from contagion channels mediated by interlinkages of financial institutions alone, but also external shocks from the real economy matter \citep{Els06}. The effects of supply chain contagion \citep{Ino19, Bar16a,Car21} are not confined to the real economy, but can be transmitted to the financial system, opening another source of systemic risk \citep{Tab24}. Financial- and supply chain network contagion channels do interact. Studying them in isolation likely leads to an underestimation of economic damages. Hence, supply chain contagion should be included in a comprehensive understanding of financial systemic risks. 

Shock propagation in production networks is traditionally studied in IO analysis. For example for the Hurricane "Katrina" event, see \citep{Hal08}. However, severe misestimates can arise when aggregating firm-level networks to sector level \citep{Die24}. The recent availability of firm-level data \citep{Pic23, Bac23} has led to a surge in research of firm-level production networks, showing the importance of individual firms and their specific position in the supply chain networks. Shock propagation along firm-level production networks can be severe, as has been shown for the 2011 earthquake in Japan \citep{Ino19, Bar16a, Car21} or the COVID-19 pandemic \citep{Pic22, Die24}. Several studies linking production and financial networks have examined the way financial shocks affect firms \citep{Hur24, Cor19, Sil18}, but the risks of contagion in production networks for the financial system are still largely unexplored. In \citep{Pol18} firms are treated as financial institutions and the connections between firms are neglected, whereas \citep{Tab24} ignores connections between financial institutions. Other attempts to combine both layers within a single framework include \citep{Kli15}, however with only one single product market or \citep{Gut20}, using only sectoral IO information. While shocks can spread in both directions between the economic and financial layers, firm-to-bank and bank-to-firm \citep{Sil18}, we focus on the firm-to-bank direction. Previous research on the coupling of financial and economic layers was limited to sectoral production networks \citep{Gut20} or limited to shocks from the financial layer \citep{Bor20a,Hur24}.

Here, we extend the model presented in \citep{Tab24} to quantify the effect of firm-level production networks on financial systemic risk. We do this by adding the possibility of interbank-loan solvency contagion between banks. The result is a two-layer model that contains both supply chain network contagion and interbank contagion. This allows us to study the relative importance of these contagion channels and, importantly, how they interact in detail. As a showcase, we use firm-level monthly employment data during the early stage of the COVID-19 epidemic in Hungary to recreate a realistic shock scenario for the firm-level supply chain network, containing 243,339 firms and 1,104,141 supplier-buyer-links.

With the two-layer shock propagation model we first examine the importance of interbank network contagion for the financial systemic riskiness of single firms in Hungary. Then we systematically shock the production network with the synthetic firm-level COVID-19 shocks to analyze the different sources of systemic risk for each individual bank. We use the risk measures Expected Loss (EL), Value at Risk (VaR), and Expected Shortfall (ES) to measure the importance of three different systemic risk channels: (i) risk directly from the synthetic COVID-19 shock scenarios, (ii) indirect risk from supply chain contagion triggered by the direct shocks, and (iii) risk from interbank contagion. We analyze how supply chain contagions amplifies the systemic risks from interbank contagion. 

\begin{figure}
    \centering
    \includegraphics[width=0.75\linewidth]{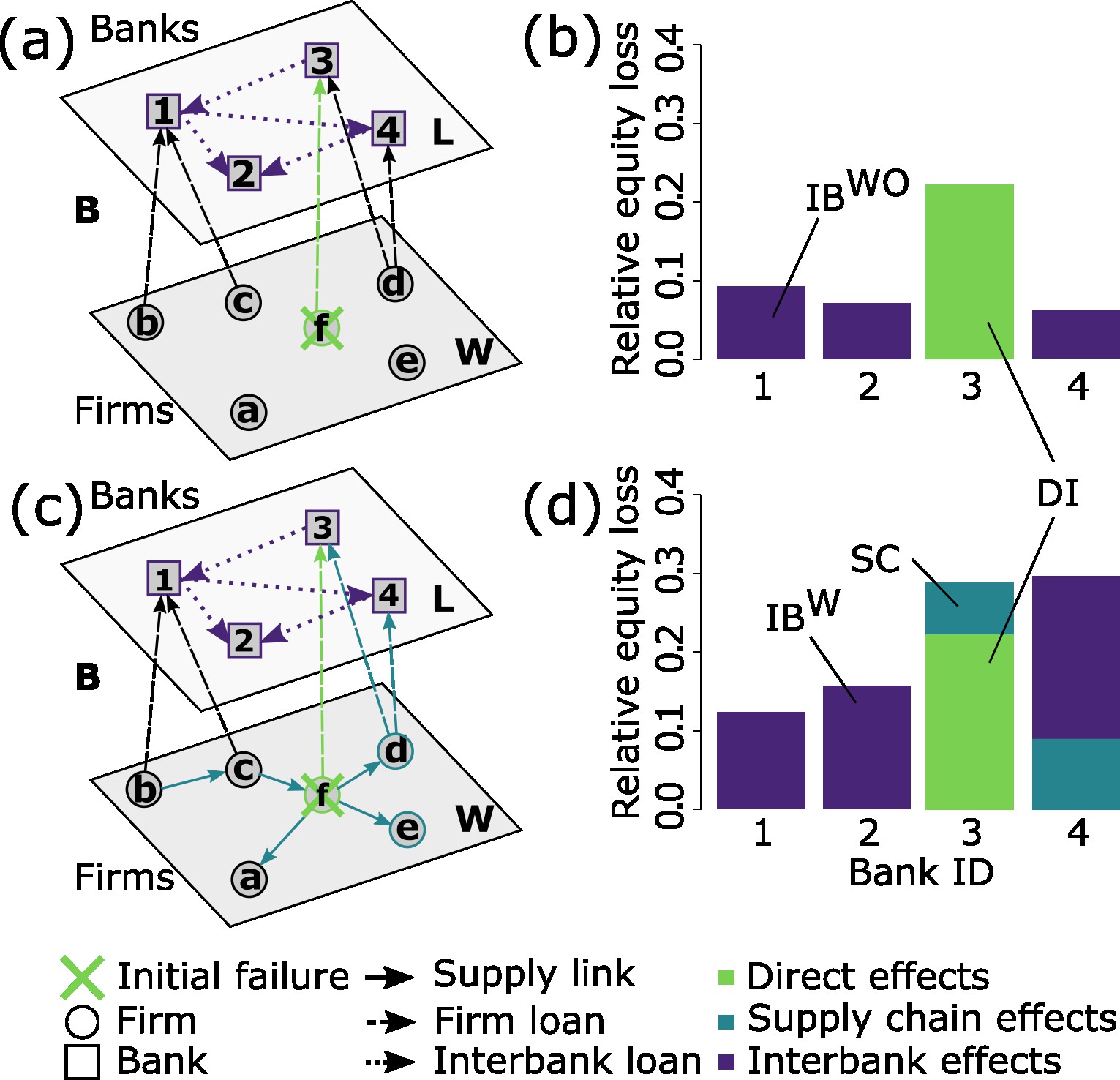}
    \caption{\textbf{Schematic depiction of the model.} (a) and (c) show the multilayer network structure with firm layer, $W$, and interbank layer, $L$. Firm loans from banks connect the two layers, marked $B$. Firms are circled letters in $W$, banks are boxed numbers in $L$. In the first scenario (a) firms are disconnected from each other, in scenario (c) they are connected via supply links (arrows). Banks are connected through interbank loans (dotted arrows). (b) and (d) show the losses for each bank in terms of the fraction of the bank's equity for scenario (a) and (c), respectively. Green color indicates losses from the initially failing firm, $f$. Light blue marks supply chain effects. Since there is no supply chain in (a) there are no supply chain effects in (b). Interbank effects are depicted in purple. The example shows the following scenario: initially firm, $f$, fails, its loans to bank 3 are written off. These direct losses are seen in (b) and (d) in the green bars. In (c) and (d) the failure of firm, $f$, causes firms, $a$, $d$, and $e$, to lose a supplier, marked by light blue arrows, which leads to a reduction in their production. This causes $d$ to default on its financial obligations as well, causing losses for banks, $3$, and ,$4$, which are represented in the light blue bar in (d). Firm, $f$, is a direct customer of $c$ and an indirect one of $d$, so they are also affected by $f$'s failure. However, they do not default on their loans and cause no further losses. After the losses from layer $W$ are propagated through $B$ to the interbank layer, $L$ (purple arrows). These respective losses are shown with purple bars in (b) and (d). $\text{IB}^\text{WO}$ stands for losses without, $\text{IB}^\text{W}$ for losses with the supply chains present.}
    \label{fig:Process schematic}
\end{figure}

\section{Model and Data}

\subsection{Model}\label{sec:Model}

\paragraph{Data}
To analyze the interplay between financial contagion and supply chain contagion we model the production network of firm interactions, interbank exposures and bank-firm loans as a multilayer network. Supply chain relations between the $n$ firms are encoded in an $n\times n$ weighted matrix, $W$, of the first layer, where the entry, $W_{ij}$, is the monetary value of the goods sold from firm, $i$, to firm, $j$, in the year 2019. The second layer is the interbank layer comprised of $m$ banks. The entry, $L_{kl}$, is the amount of money bank, $k$, borrowed from bank, $l$, ($l$ lends to $k$). We use indices $k,l\in \left\lbrace1,2,...,m\right\rbrace$ to refer to banks and $i,j \in \left\lbrace1,2,...,n\right\rbrace$ for firms. Since a loan needs to be repaid, this is a liability for bank, $k$, and an asset for bank, $l$. The two layers are connected through an $n\times m$ matrix, $B$, where the entry, $B_{ik}$, denotes the amount of outstanding principle of the loans bank, $k$, lent to firm, $i$, which equals the liabilities of firm, $i$, to bank, $k$. $B_{ik}=500 \text{ HFN}$ means that bank, $k$, would lose $500\text{ HFN}$ of its equity, $e_k$, should firm, $i$, default on its loan. We calculate the losses of a bank relative to its equity. In this example, if bank, $k$, had an equity of $1,000\text{ HFN}$, that would translate into losing ${B_{ik}}/{e_k}=50\%$ of its equity should firm, $i$, fail to repay its loan. The same applies to interbank links, $L_{kl}$.

\paragraph{Model summary} 
To estimate the relevance of supply chain contagion on financial systemic risk, we compare two different multilayer networks, shown schematically in Fig.~\ref{fig:Process schematic}~(a) and (c). The lower layer, denoted by $W$, represents the buyer-supplier relations between the 6 firms, $a, \cdots, f$. Supply chain relations are represented by arrows. In (a) all $W_{ij}$ are set to 0, whereas in (c) we take supply chains into account, e.g. firm, $b$, delivers goods to firm, $c$, i.e. $W_{bc}\neq 0$. The interbank layer is denoted by $L$ and is the same in both cases. $B$ represents the financial connection from the firm layer, $W$, to the interbank layer; it stays the same too. This comparison allows us to examine the interaction between the production network and interbank network contagion channels and their relative importance in creating financial systemic risk. We use the following three steps in the corresponding simulations.
\begin{enumerate}
    \item apply an exogenous shock to the production level of individual firms, which then can propagate through the supply chain,
    \item examine the financial health of firms using their financial statements and compute the losses for all individual banks,
    \item the resulting defaults on loans are then used as an exogenous shock on banks assets which can propagate financial distress through the interbank layer.
\end{enumerate}
We reconstruct an agent-specific 1:1 representation of the Hungarian economy as it has occurred in the year 2019 based on detailed information on balance sheets, supply relations, bank- and interbank loans. 

\paragraph{Shock propagation in the supply chain}
We use a supply chain propagation mechanism developed in \citep{Die22} and used in \citep{Rei22,Die24,Die24a,Sta24,Fes24}. The mechanism produces an estimate for the economic systemic risk, ESRI, a systemic risk value attached to every firm, $i$, that indicates the reduction of the total system output in case $i$ would exit the supply chain. 
A key quantity in the ESRI algorithm is the remaining production level of $i$ , $h_i(T)$, after $T$ contagion iterations after an initial shock, $h_i(t_1)=\psi_i\in\left[0,1\right]^n$. The model uses Generalized Leontief production functions (GLPF) calibrated from the supply chain network data, the firms' industry sectors, and a list of essential and non-essential inputs for production (derived from a sector survey \citep{Pic22}). 
For implementing exogenous shocks, we use the following notation for a shock applied to firm, $i$. $\psi_i=1$ means that it produces according to its production function, $\psi_i=0$ means it does not produce anymore due to a 100\% shock. 
In the examples in Fig.~\ref{fig:Process schematic}~(a) and (c) production of firm, $f$, is initially reduced to $0$\% of its production, using $\psi_f = 0$ and $\psi_i = 1$ for $i \neq f$. The reduction of the production level of firm, $f$, leads to less goods delivered to its customers, firms $a$, $d$, and $e$, and subsequently to a reduction of their production levels as well. This downstream contagion of production losses is referred to as a supply shock. On the other hand, firm, $f$, also stops purchasing inputs from its suppliers, here, firm $c$. These firms then reduce their production output to meet the decreased demand. We refer to this as upstream contagion or as a demand shock. Using the updated production levels, $h_i(t_2)$, the model continues iterating and propagates the initial shock up- and downstream along the supply chain. The model converges to a new stable state at time $T$ once $h_i(T-1)-h_i(T)\leq\epsilon$ for all $i\in\{1,...,n\}$, where we set $\epsilon = 0.01$. Finally, the vector of remaining production levels is $h_i(T)\in\left[0,1\right]^n$. $h_i(T)=0.3$ means that firm, $i$, loses $70$\% of its production after the shock event propagated through the entire supply chain. For the network without supply chain links in Fig.~\ref{fig:Process schematic}~(a) there is no such contagion in the supply chain and $h_i^\text{WO}(T)=h_i(t_1)=\psi_i$, where the superscript $\text{WO}$ denotes the case without supply chain contagion. To distinguish the two model specifications, we call the case with supply chain contagion $h_i^\text{W}(T)\in\left[0,1\right]^n$. In the example with firm, $f$, initially failing, this translates into $h^\text{WO}(T)=\left(1,1,1,1,1,0\right)$ and $h^\text{W}(T)=\left(0,0,0,0,0,0\right)$. We refer to this process as supply chain contagion or supply chain shock propagation.

\paragraph{Linking to the bank network} In a second step we now calculate losses of banks from the firms' remaining production levels, $h_i(T)$, using the financial statements of firms and the loan portfolio data of every bank as in \citep{Tab24}. From a firm's financial statements we know its revenue, $r_i$, and its operating costs, $c_i$, which (along with other income statement items) make up its profit and which are affected by the production level reduction. A reduction of a firm's production level decreases both, allowing us to calculate the change in profit of firm, $i$, after the initial shock and the propagation as
\begin{equation}
    \Delta p_i^\text{W} = \left(1-h_i^\text{W}(T)\right)\left(r_i-c_i\right)\quad  \label{eq:DeltaProfit}
\end{equation}
and, analogously, $\Delta p_i^\text{WO}$, for $h_i^\text{W}$. This difference in profit is the difference between operating revenue, $r_k$, and cost, $c_i$, multiplied by the reduction in production level, $1-h_i(T)$. To calculate the losses of banks we need to determine if the financial losses (profit reduction from the counterfactual shock) are big enough for the firm to default. The profit loss can be interpreted as the difference between the counterfactual scenarios with shocks and the unshocked world, as represented by the data. For this we define the default indicator of firm, $i$, for the supply chain contagion specification as
\begin{equation}
    \chi_i^\text{W}(\psi)=
    \begin{cases}
        1 \text{ if } (z_i-\Delta p_i^\text{W})\leq 0 \text{ or } (a_i-s_i-\Delta p_i)\leq 0 \quad , \\
        0 \text{ if } (z_i-\Delta p_i^\text{W})> 0 \text{ and } (a_i-s_i-\Delta p_i)> 0\quad .
    \end{cases}\label{eq:Chi}
\end{equation}
Analogously, for the case without supply chain contagion, $\chi_i^\text{WO}$ is given by $\Delta p_i^\text{WO}$. Here $z_i$ denotes the equity of firm, $i$, $a_i$ denotes its short-term assets and $s_i$ are short-term liabilities. A firm defaults, $\chi_i=1$, if its profit losses are larger than either its equity, $z_i$, the first part of the conditional, or its available short-term liquidity, $a_i - s_i$, the second part of the conditional. If a firm fails we assume that all outstanding principle of its loans is written off by the bank. For simplicity, we assume that the \textit{loss given default (LGD)} is $100\%$ for all loans. It was shown in \citep{Tab24} that for any other choice of LGD results just scale proportionally as long as banks don't fail. Here we can distinguish between two sources of losses for banks, the direct losses of bank, $k$, caused by the initial shock to the firms, $\text{DI}_k$, and the additional losses caused by the supply chain contagion, $\text{SC}_k$. The losses of bank, $k$, are given by
\begin{align}
    \text{DI}_k &= \sum_{i=1}^n \chi _{i}^\text{WO} \frac{B_{ik}}{e_k} \label{eq:Direct}\quad ,\\
    \text{SC}_k &= \sum_{i=1}^n \left(\chi _{i}^\text{W}-\chi _{i}^\text{WO}\right) \frac{B_{ik}}{e_k}\quad ,
    \label{eq:SC} 
\end{align}
where $B_{ik}$ is the outstanding principle of loans firm, $i$, received from bank, $k$, normalized by the Tier 1 Equity, $e_k$, of bank, $k$. $\chi_i^\text{WO}$ and $\chi_i^\text{W}$ are the default indicators defined in Eq.~(\ref{eq:Chi}). The term $\chi _{i}^\text{W}-\chi _{i}^\text{WO}=1$ only if a firm defaults on its loans when supply chain contagion is taken into account, but not without that mechanism (i.e., not from the initial shock). Figure~\ref{fig:Process schematic}~(b) and (d) show this more clearly. The bars show the relative amount of equity lost for each of the four banks and the color denotes the source of the loss. Panel (b) corresponds to the case with no supply chain contagion. Firm, $f$, fails initially and defaults on its loans. This directly causes bank, $3$, to lose $25$\% of its equity, marked by the green bar. When considering supply chain contagion firms, $e$, and, $d$, also default, while the other firms have enough equity and liquidity buffers to absorb the losses. Firm, $e$, has no loans and causes no losses to any bank. Firm, $d$, on the other hand, has taken loans from banks, $3$, and, $4$, which are written off and lead to additional supply chain contagion induced losses. Bank, $3$, loses an additional $\text{SC}_3=5\%$ of its equity and bank, $4$, loses $\text{SC}_4=10$\%. The total losses of bank, $k$, in terms of percent of equity lost at this point is given by $\mathcal{L}(\psi,t_1)^\text{W}_k=\text{DI}_k+\text{SC}_k$ and $\mathcal{L}(\psi,t_1)^\text{WO}_k=\text{DI}_k$.

\paragraph{Interbank contagion with DebtRank} In a third step the resulting bank losses from the initial shock and its propagation are propagated over the interbank layer, $L$, using the 2015 adaptation \citep{Bar15} of the DebtRank\citep{Bat12} algorithm. In short, if a bank, $l$, has extended a loan to bank, $k$, i.e. $L_{kl}\neq0$. This loan is a liability on the balance sheet of bank, $k$, but an asset for $l$. The idea behind the DebtRank algorithm is that if $k$ loses a certain percentage of its equity then the value of the loan, $L_{kl}$, an asset on the side of bank, $l$, loses the same fraction. This credit value adjustment mechanism gives rise to an iterative process where accounting losses from value adjustments are spread from bank to bank.\footnote{This is the underlying idea of mark-to-market accounting.} At iteration step, $t$, the total Losses of bank, $k$, (including the direct shock, supply contagion induced bank-firm loan write offs and, interbank solvency contagion) can be calculated as
\begin{equation}
    \mathcal{L}^\text{W}_k(t,\psi) = \mathcal{L}^\text{W}_k(t-1,\psi)+\sum_{l=1}^m \Lambda_{lk} \,\min{\left(\mathcal{L}^\text{W}_l(t-1,\psi),1\right)} \quad . \label{eq:DR}
\end{equation}
The case without supply chain contagion works in the same way, using $\mathcal{L}^\text{WO}_k(t,\psi)$. $\Lambda_{lk}=\frac{L_{lk}}{e_k}$ is called the interbank leverage matrix. In the original DebtRank \citep{Bat12} a bank only spread a shock to other banks once. 
We use a linearized version of DebtRank \citep{Bar15}, where banks can spread shocks until they default or the algorithm converges. In Eq.~\ref{eq:DR} the minimum function limits the losses any individual bank can cause in any other bank to the size of the loan. The DebtRank algorithm is initiated at time $t_1=1$ with $\mathcal{L}^\text{W}_k(t_1,\psi) = \text{DI}_k+\text{SC}_k$ and $\mathcal{L}^\text{WO}_k(t_1,\psi) = \text{DI}_k$. The process terminates at time $T$, once the change of equity in the banking system is smaller than $\epsilon$, i.e. when $\frac{\sum_k\mathcal{L}_k\left(t,\psi\right) - \mathcal{L}_k\left(t-1,\psi\right)}{\sum_k e_k}\leq \epsilon$ with $\epsilon=0.01$. In Fig.~\ref{fig:Process schematic}~(b) and (d) the additional losses due to interbank contagion, $\text{IB}^\text{W/WO}_k=\mathcal{L}_k^\text{W/WO}(T,\psi)-\mathcal{L}_k^\text{W/WO}(t_1,\psi)$, are shown with purple bars. Additional interbank losses in the presence of supply chain contagion are always equal to or larger than without supply chain contagion $\text{IB}_k^\text{W}\geq \text{IB}_k^\text{WO}$, since we only consider negative shocks to production.

\paragraph{Impact of single firm failures} In the first part of our analysis we estimate the impact of single firms' hypothetical failures on banks' equity. We define the single-firm shock vectors as
\begin{equation}
    \psi_i^j = \begin{cases}
        0\text{ for }i=j \quad , \\
        1\text{ otherwise} \quad .  \label{eq:Single_firm_shock}
    \end{cases}
\end{equation}
As in the example in Fig.~\ref{fig:Process schematic}, we set firm $f$'s production capacity to $0$. Following the definition in \citep{Tab24} we calculate the financial systemic risk index, FSRI, of each firm by summing the total amount of equity lost in the banking system after propagating the initial disruption through the supply chain network and calculating the defaults of firms (not considering interbank contagion),
\begin{equation}
    \text{FSRI}_i = \sum_{k=1}^m \frac{e_k}{\sum_{l=1}^m e_l}\min\left[\mathcal{L}_k(1,\psi^i),1\right] \quad . \label{eq:FSRI}
\end{equation}
Here, the sums extend over all banks, $k$, or $l$, and $e_k$ is the bank's equity. $\mathcal{L}_k(1,\psi^i)$ denotes the relative amount of equity bank, $k$, loses after firm, $i$, shuts down and the disruption is propagated through the supply chain. Note that FSRI does not take the interbank market into account. We thus generalize the measure by accounting for interbank solvency contagion and define
\begin{equation}
    \text{FSRI}^+_i = \sum_{k=1}^m \frac{e_k}{\sum_{l=1}^m e_l}\min\left[\mathcal{L}_k(T,\psi^i),1\right] \quad . \label{eq:FSRI+}
\end{equation}
Note the difference to FSRI as here we use $\mathcal{L}_k(T,\psi^i)$, i.e., banks losses after the initial shock propagates via the supply chain, causes losses in the banking sector \emph{and} those then propagate through the interbank market -- where we use the normal DebtRank mechanism.

\paragraph{Shocks scenarios including multiple firms}
To investigate the interplay between the interbank and the supply chain layer we want to use more general and realistic firm shocks, $\psi$, e.g., derived from employment data during the early stages of the COVID-19 pandemic. The way we proceed is described in detail in \citep{Die24}, here, we give just a brief overview. We have access to the number of employees of 160.000 firms in Hungary. By assuming that labor is a Leontief-type input for the production process, we use the relative reduction in employment from January to May of 2020 to infer a likely production shock to all the firms. For firms with missing employment data we impute shocks by drawing production level reductions from the available firms in the same NACE 4-digit industry. There were no furlough schemes in Hungary, which allows us to proxy the production level reductions via this employment data.\footnote{This does not mean that we correctly capture the actual reduction of production levels during the first quarter of 2020. Rather, this so-constructed shock scenario should be seen as one possible, however, realistic approximation of the firm-level production level reduction.} To create a distribution of shocks with realistic magnitudes we use this empirical result to create 1,000 synthetic independent statistically identical COVID-19 shock scenarios. We do so by randomly drawing shocks for specific firms, while rescaling them to guarantee that the shock is the same as the empirical one when aggregated to the NACE 2-digit industry level. The result is a set of 1,000 artificial shocks that differ on the firm-level but are the same as the (typically) empirically observed one, when aggregated to the industry-level. It has been shown that the resulting firm-level shocks coupled with our shock propagation model for the production network result in a percentage reduction of total network gross output that is indeed similar to the observed reduction in GDP of Hungary during the first quarter of 2020 \citep{Die24}. 

To analyze the financial risks of the resulting distribution of equity losses of individual banks across the 1,000 synthetic shocks, we use the standard measures of expected loss (EL), defined as the mean of the distribution, value at risk (VaR), the 95th quantile, and the expected shortfall (ES), the average of all losses in the top 5\% quantile. 

\subsection{Data}

\paragraph{Overview} 
We use real-world data to calibrate the presented model. In particular, we use the following datasets available at the Central Bank of Hungary,
\begin{itemize}
    \item value added tax (VAT) reports from 2019 for all firms in Hungary reporting firm-firm VAT above a threshold of 100,000 HUF, to proxy $W$,
    \item corporate tax reports of firms and corporate registry data, for $z_i, a_i, s_i$, 
    \item credit and bank equity information compiled for the financial stability report of the Central Bank for $B$, $L$, and $e_k$.
\end{itemize}

\paragraph{Value added tax data} 
Firms' value added tax reports are used to reconstruct the supply chain network, $W$. We use data from the year 2019 that contains records of all transactions among the 243,339 firms that exceeded a tax content threshold of 100,000 HUF (approximately 250 EUR). To filter incidental transactions we retain only stable supply links, defined as those where transactions occur in at least two different quarters in 2019. This removes 48\% of recorded links, while keeping 93\% of the national transaction volume. For more details, see \citep{Bor20}. Further information on the characteristics of the resulting network are found in \citep{Die22} and \citep{Bac23}. This dataset also contains information on the employment numbers, which allows us to use the observed employment reduction from January to May of 2020 to construct realizations of realistic shock scenarios modeled after the actual COVID-19 crisis. For details, see \citep{Die24}.

\paragraph{Financial statements of firms}
The date on the income statements and balance sheets of firms (obtained from corporate tax data) is used to calculate the parameters in Eqs.~(\ref{eq:DeltaProfit}) and (\ref{eq:Chi}), in particular, revenue, costs, equity, and short-term assets and liabilities necessary to calculate liquidity. Because different economic entities are covered by the corporate and VAT tax reports, income data of 65,606 of the 243,339 firms in the network is missing. Additionally, we exclude 36,501 firms due to negative equity, liquidity, or net income in their financial statements. Firms with negative equity or liquidity would be classified as defaulted by our definition, see Eq.~(\ref{eq:Chi}), and firms with negative net income would experience increased equity with reduced operational activity. Although these firms cannot directly cause losses in the financial system by defaulting on loans, they are still included in the supply chain contagion. This allows them to indirectly generate losses in the financial layer through supply chain contagion. For further details on the data, see \citep{Tab24}. 

\paragraph{Bank and credit information}
We use the credit registry to calibrate the model by reconstructing the matrices, $B$ and $L$. The credit registry contains information on different kinds of contracts, e.g., leases. We filter the dataset for loan contracts and remain with data on the loan exposures of 40,043 firms to 269 banks in Hungary. The remaining companies do not have reported bank loans. For the banks we use the same dataset as in the stress testing framework of the financial stability report of the Central Bank of Hungary \citep{Mag24}, which contains information on 19 banks and their Tier 1 Equity. We omit the following types of institutions: local institutions that are not relevant for bank supervisory purposes as well as institutions with a large share of non market based operations, such as infrastructural development institutions. The resulting firm-bank loan matrix, $B$, is of size $n\times m$ with $n=243,339$ and $m=19$. In this way we cover 60.7\% of all outstanding firm-bank loans. The interbank network, $L$, is reconstructed from short-term unsecured loans between banks. The short maturity of these loan means a single snapshot is not representative of the actual interdependencies of banks. Thus we construct the network by calculating the highest daily exposure of each bank during the fourth quarter of 2019, which is then distributed in proportion of the average exposures towards the banks' partners in the same time frame. 

With these datasets the model is fully determined and entirely data-driven. The remaining modeling choices are limited to the mechanism of the supply chain contagion, see \citep{Die22}, assumptions on the effect of production shocks on a firms' financial health \citep{Tab24}, and the contagion mechanism in the interbank layer \citep{Bar15}. In particular, we assume a reduction in revenue and material costs being proportional to the reduced production. Further, we assume a loss-given-default of 100\%.

\section{Results}

We analyze the importance of the three contagion channels --- direct shocks (DI), supply chain contagion (SC), and interbank contagion (IB) for amplifying the financial systemic risk that firms create for banks and the entire system. We simulate a shock scenario to the supply chain network and sequentially open the supply chain- and interbank contagion channels. 
For every contagion channel, the simulations yield the bank equity losses for the particular shock scenario. DI$_k(\psi)$, SC$_k(\psi)$, and IB$_k(\psi)$ denote the equity loss of bank, $k$, for shock scenario, $\psi$, for the three channels respectively. We define, 
\begin{eqnarray}
\text{DI}(\psi) \equiv\sum_k \frac{e_K}{\sum_l e_l}\text{DI}_k(\psi) \quad    \\ 
\text{SC}(\psi) \equiv  \sum_k \frac{e_K}{\sum_l e_l}\text{SC}_k(\psi) \quad    \\ 
\text{IB}(\psi) \equiv \sum_k \frac{e_K}{\sum_l e_l}\text{IB}_k(\psi) \quad .  
\end{eqnarray}

\subsection{Contagion channel relevance for financial systemic risk of individual firms}\label{sec:FSRI}

\begin{figure}[t]
    \centering
    \includegraphics[width=\linewidth]{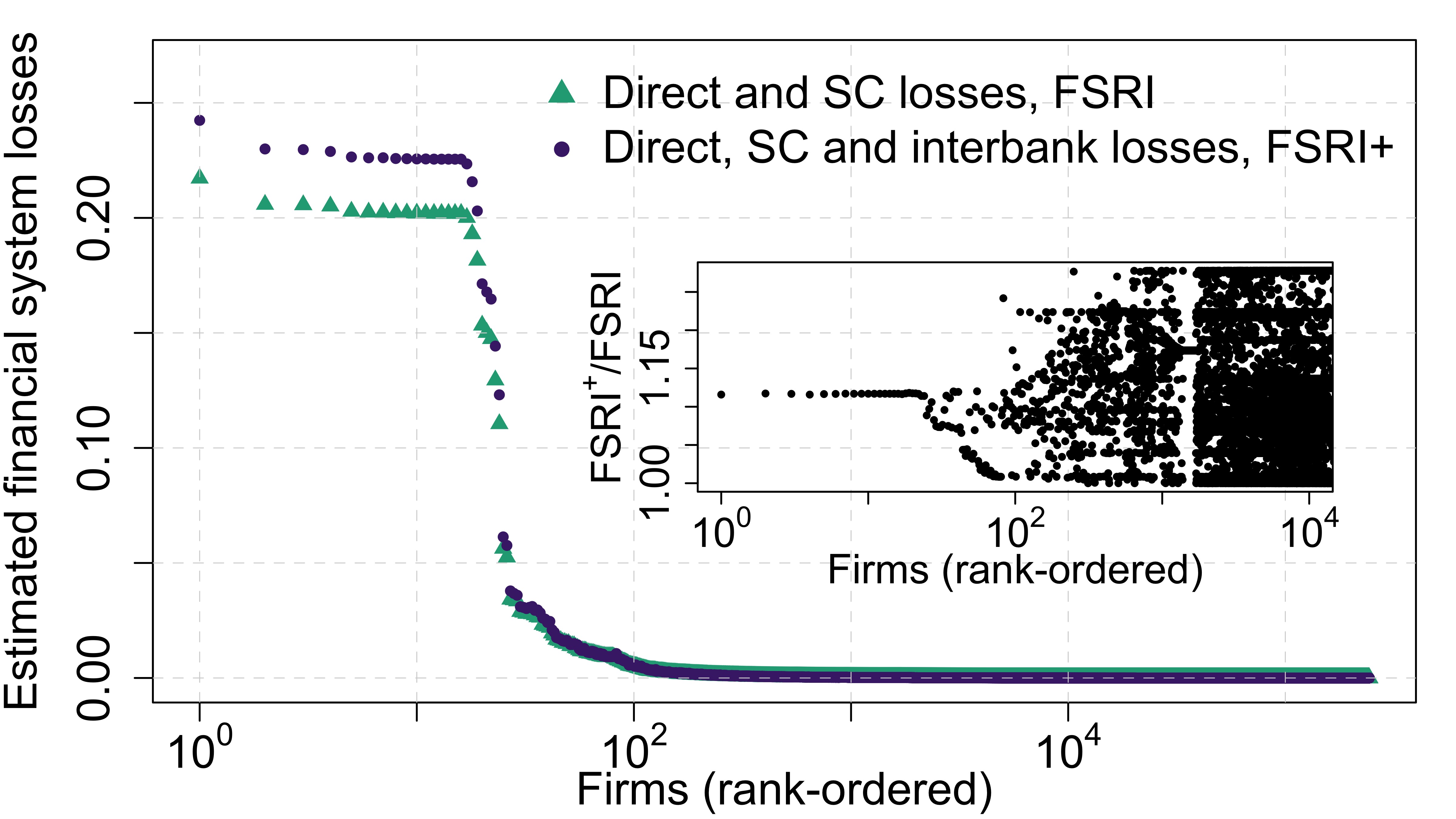}
    \caption{\textbf{Systemic risk profiles:} Financial Systemic Risk Index, FSRI (light blue), and Financial Systemic Risk Index with interbank contagion, FSRI+ (purple), for every firm in the Hungarian production network (x-axis, rank ordered). The impact of the failure of every firm on the production network is computed with the ESRI algorithm. The resulting bank losses are calculated and the sum of lost equities of banks is presented as a fraction of the total equity in the banking sector, see Eq.~\ref{eq:FSRI}. Next, the bank-wise losses are propagated on the interbank market, resulting in higher losses, see Eq.~\ref{eq:FSRI+}. The inset shows the relative financial risk increase for all firms when the interbank contagion is taken into account. The firms in the plateau can cause 12\% higher losses if the interbank contagion channel is taken into account. For the vast majority of firms risk increases between 0 and 28\%.}
    \label{fig:FSRI+}
\end{figure}

First, we illustrate how the interbank network amplifies the financial systemic risk of individual firms. We assume the hypothetical failure of a single firm, as defined in Eq.~(\ref{eq:Single_firm_shock}), and calculate the losses for banks after the failure triggers cascading events in the supply chain network. This cascading leads to additional defaults on loans and financial losses for banks. Summing the total losses for all banks, relative to the equity of the financial system, we estimate the FSRI for every firm, $i$. 
It quantifies the effect that the failure of a single firm poses to the financial system. When we include the interbank contagion channel, we use, $\text{FSRI}^+_i$, from Eq.~(\ref{eq:FSRI+}).
Figure~\ref{fig:FSRI+} compares the rank ordered raw FSRI (light blue) with the interbank-contagion-augmented $\text{FSRI}^+$ (purple). The most risky firms appear to the very left. The plateau on the left is due to a small subset of 19 firms affecting more than 20\% of the banking system's equity and 24 firms affecting more than 10\%, see also \citep{Tab24}. Without the additional interbank contagion, the most risky firms can affect up to 22\% of the total equity in the financial system. This increases to 24\% when including the possibility of interbank contagion. 
The inset shows the relative increase of $\text{FSRI}^+$ compared to FSRI. The systemic risk index of the plateau firms increased by 12\%, an amplification factor of $1.12$, due to the presence of the interbank network.
This similarity in relative increase is explained by the fact that firms in the plateau form a 'systemic risk core' in the supply chain network, where any failing firm in that core leads to the failure of all the others in the core, see \citep{Die22}, and see \citep{Tab24} for how this translates to the equity losses of banks. The observed amplification of a factor $1.12$ is driven by the interbank network volume being substantially smaller than the total exposure of firms to the supply chain network by an order of magnitude, i.e., $\frac{\sum_{ik}B_{ik}}{\sum_{kl}L_{kl}}=12.5$. Looking at individual banks the median of total exposure to firms relative to total exposure to banks, i.e. $\frac{\sum_{i}B_{ik}}{\sum_{k}L_{kl}}$, is $10.5$. In countries with larger interbank networks, or when considering additional exposure channels as, e.g., in \citep{Pol15}, amplification from IB contagion could be substantially larger. 

Note that the relative bank equity loss increase of firms not part of the plateau is heterogeneously distributed. Next we explain how the horizontal stripes in the inset of Fig.~\ref{fig:FSRI+} emerge. Consider two firms, $i$ and $j$, causing a supply chain cascade that results in losses only to a \emph{single} bank, $k$. In both cases the interbank contagion is triggered only by the equity loss of bank, $k$, and the resulting losses are proportional to the DebtRank of bank, $k$, (corresponding to the full failure of bank, $k$). Note that the interbank contagion is linear, as the interbank layer is relatively small. Hence, for both firms, $i$ and $j$, the initial losses caused to the banking layer are amplified by the same linear factor, yielding the horizontal stripes. If that bank, $k$, has no liabilities, i.e., $\Lambda_{kl}=0$ for all $l$, then there is no additional cascade in the financial layer and the amplification factor $\frac{\text{FSRI+}_i}{\text{FSRI}_i}=1$. If bank, $k$, has loans, the amplification factor can be as high as $1.28$ for cases when $k$ has a high debt-to-equity ratio, $\frac{\sum_l L_{kl}}{e_k}$. Here, $L_{kl}$ is the amount of money lent from bank $j$ to bank $k$ with equity $e_k$, see also Sec.~\ref{sec:App_FSRIAmp} for a more detailed derivation. This agrees well with previous findings in \citep{Die20}, where it was shown that DebtRank tends to increase for banks with higher liabilities.

The structure of the production network leads to a fat-tailed distribution of financial systemic risk, shown in Appendix Fig.~\ref{fig:Fig2_Survival}. Even without the plateau firms, the distribution of financial losses decays slowly and 72 firms can cause losses of up to 1\%, see also \ref{sec:App_FSRIDistri}. When compared to DebtRank, shown in Fig.~\ref{fig:Fig2_DebtRanks}, the effect of plateau firms is higher than the default of the second largest bank, which leads to losses of 16.6\% of equity in the financial system. In both cases the effect of the interbank contagion is relatively small when compared to the initial shock of the interbank layer.

Losses from interbank contagion are small compared to the effects of supply chain contagion, but they are by no means negligible. Even though the interbank layer is relatively small, it can still inflate financial systemic risk of firms by about 10\%. Even for firms causing small supply chain network cascades, bank equity losses are amplified through interbank contagion.

\subsection{Contagion channel relevance for bank equity losses from COVID-19 shocks}

Next, we analyze the effect of direct losses, supply chain contagion, and interbank contagion on financial losses of banks by using the 1,000 synthetic COVID-19 shocks. Every individual shock constitutes one possible realization of the empirical COVID-19 shock during the first quarter of 2020. For details, see Sec.~\ref{sec:Model} and \citep{Die24}. 

\begin{figure}[t]
    \centering
    \includegraphics[width=\linewidth]{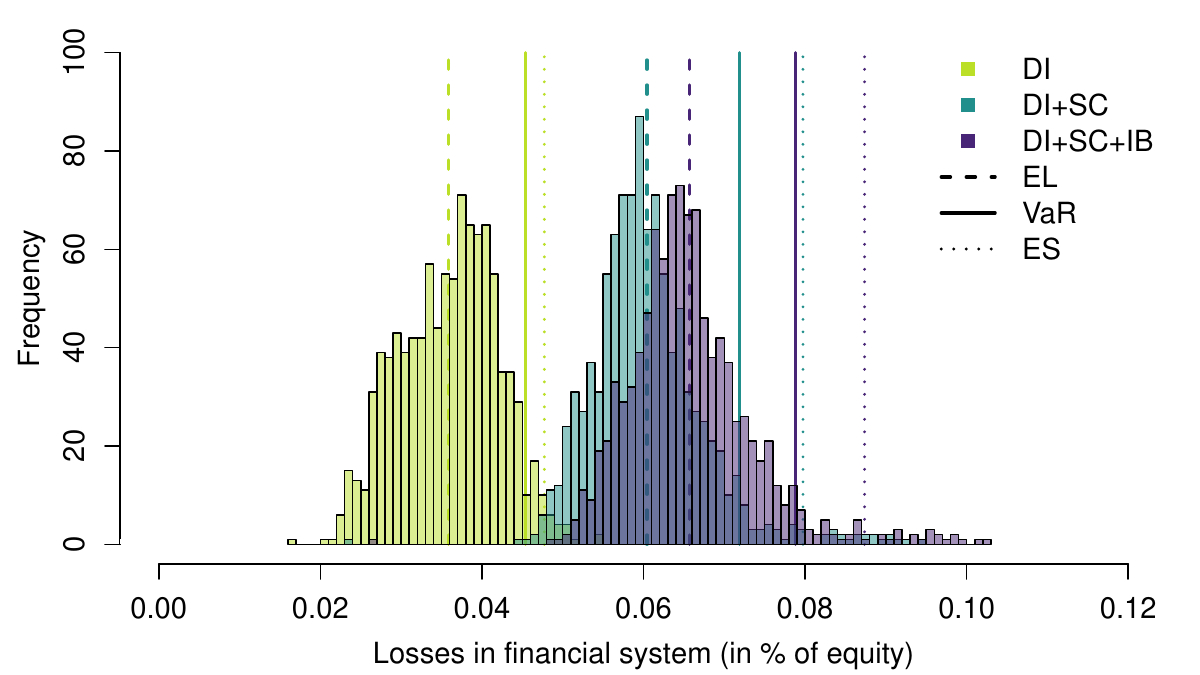}
    \caption{\textbf{Equity loss distribution for the banking system, for 1,000 synthetic COVID-19 type shocks for the three contagion channels.} Green: Direct losses from firm failures caused by the initial shock, blue: With additional losses from supply chain contagion, Purple: With additional losses from interbank contagion. The vertical lines denote the three different risk measures, Expected Loss (dashed), Value at Risk (solid), and Expected Shortfall (dotted).}
    \label{fig:SystemLossHist}
\end{figure}

\subsubsection{Contagion channel relevance for banking system losses}

Figure~\ref{fig:SystemLossHist} shows the distribution of aggregated bank system equity losses for the 1,000 synthetic COVID-19 shock scenarios after each of the 3 contagion channels (sequential simulation steps). Direct losses, DI, (green) range from 1.6\% to 5.5\%; losses with supply chain contagion, DI+SC, (blue) range from 2.4\% to 9.4\% and losses with additional interbank contagion, DI+SC+IB (purple) range from 2.6\% to 10.3\%. Note that DI, SC, and, IB are vectors of length 1,000, each element corresponds to the system equity loss for one of the 1,000 COVID-19 shock scenarios. A pairwise two-sample Welch test between the three distributions confirms that the mean values are different with a $p$-value less than $10^{-15}$.
While DI is slightly left skewed, the losses with supply chain contagion (DI+SC) show a fat tail towards large losses. Including interbank contagion (DI+SC+IB) the tail becomes slightly more pronounced. This is evident by looking at the difference between ES and VaR for each of the distributions. For DI+SC the difference is ES(DI+SC)-VaR(DI+SC)=0.78\% and increases to 0.85\% for DI+SC+IB. For a reference, for DI alone, the difference is 0.23\%. The three financial risk measures are seen in Fig.~\ref{fig:SystemLossHist} as vertical lines, Expected Loss (dashed), Value at Risk (solid) and Expected Shortfall (dotted). EL is found to be 3.6, 6.0, 6.6\% for DI, DI+SC, and DI+SC+IB, respectively, while VaR is 4.5, 7.2, 7.9\%, and ES is 4.8, 8.0, 8.7\%. 
The average direct loss, EL(DI), is amplified by 68\% when taking into account supply chain contagion, (EL(DI+SC)/EL(DI)) and 83\% for (EL(DI+SC+IB)/EL(DI)). The tail of direct losses, ES(DI), is amplified by 67\% when taking into account supply chain contagion (ES(DI+SC)/ES(DI)) and 83\% for (ES(DI+SC+IB)/ES(DI)). 
This means that the direct shock channel contributes more than half of the overall losses in the banking system. While supply chain contagion still contributes substantially, interbank losses are again relatively small, however, not negligible. For more details, see Sec.~\ref{sec:App_bankwiserisk}. 

\begin{figure}[t]
    \centering
    \includegraphics[width=\linewidth]{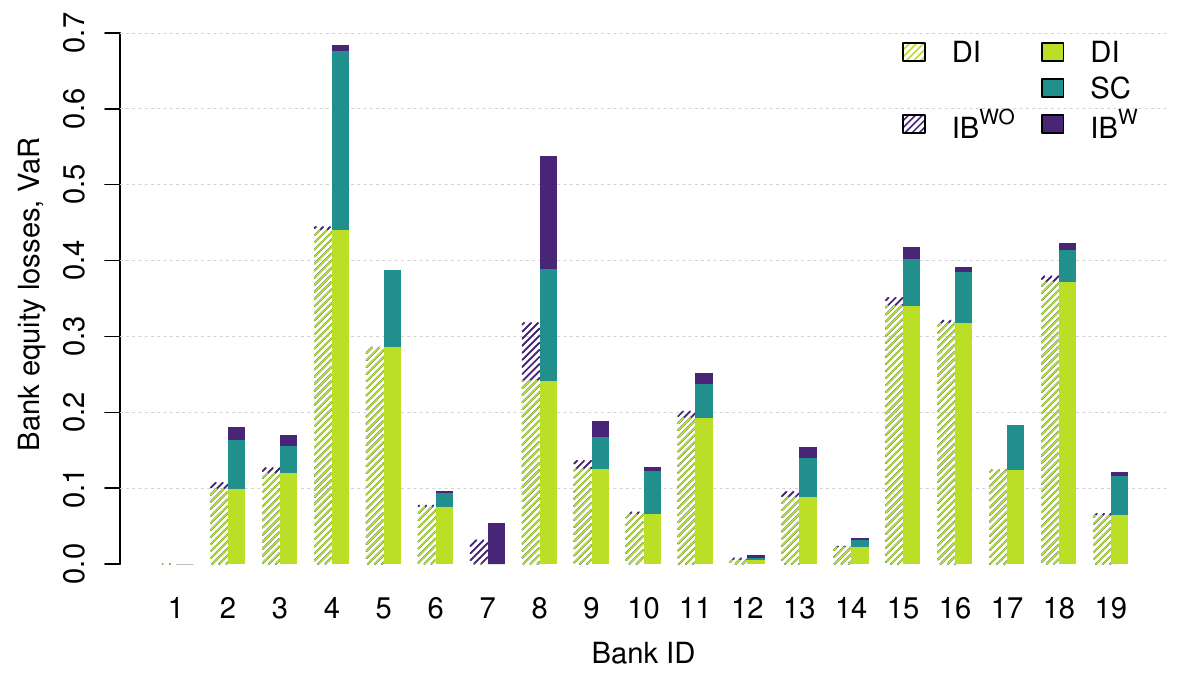}
    \caption{\textbf{Value at Risk (VaR) at the 95\% level from the loss distribution created by 1,000 artificial COVID-type scenarios for all 19 banks.} The solid and shaded bars correspond to the scenario with and without supply chain contagion, respectively. Colors indicate the effects from the different parts of the model: Green,losses from the initially failing firm, $\text{DI}$; Blue marks supply chain effects, $\text{SC}$. Since there is no supply chain in the first scenario there are no shaded blue bars. Losses from the interbank contagion are in purple $\text{IB}^\text{W/WO}$. One can see that for most banks, losses from the supply chain, either direct, or indirect are the most relevant. Some banks, 7 and 8 specifically, are exposed to non-trivial amounts of risk from the interbank market. The total amount of risk exposure varies for each bank from 0 to 70\% of the bank's equity.
    }
    \label{fig:Fig_3}
\end{figure}

\subsubsection{Contagion channel relevance for the risk of individual banks}

We now assess the importance of the three contagion channels for \emph{individual} bank equity losses. We focus on VaR; for EL and ES see ~\ref{sec:App_RiskMeasures}. 
Figure~\ref{fig:Fig_3} shows the Value at Risk (VaR) at the 95\% level for all banks. 
Colors indicate the contagion channels, Green is the direct loss channel, blue represents losses from supply chain contagion, and purple bars indicate losses due to interbank contagion. 
To see the full importance of the supply chain contagion channel, we display two versions of VaR. The first is VaR(DI$_k$+SC$_k$+IB$_k$), as before (solid colored bar). For the second version we apply interbank contagion directly after the direct impacts, DI, (hatched colored bar), i.e., without considering supply chain contagion. This is necessary as the presence of supply chain contagion can affect the magnitude of interbank contagion. The difference in bar height is the total increase of banks' VaR from supply chain contagion.

The VaR of equity losses without supply chain contagion, VaR(DI$_k$ + IB$_k^\text{WO}$), (hatched bars) of individual banks, $k$, with the first quartile at 6.7\% and the third a 30.2\% of banks' equity, with a median value of 12.4\%. The smallest VaR(DI$_k$ + IB$_k^{WO}$), excluding bank 1 with no losses, is 0.8\% and the largest is 44.4\%. Supply chain contagion (solid bars) increases the VaR of individuals banks; the first and third quartile of the banks, VaR(DI$_k$+SC$_k$+IB$_k$), are at 10.9\% and 38.9\%, respectively, with the median at 18.1\%. The minimum and maximum values of VaR(DI$_k$+SC$_k$+IB$_k$) are 1.2\% and 65.5\%. The resulting amplification factors for the total VaR of individual banks' equity losses, VaR(DI$_k$+SC$_k$+IB$_k$) / VaR(DI$_k$ + IB$_k^{WO}$), have their first quartile at 1.27 and the third at 1.67 with a median at 1.48. The smallest amplification factor is 1.12, the largest is 1.87.
Supply chain contagion amplifies VaR with respect to direct losses of banks $k$ (not considering IB effects). For $\text{VaR}(\text{DI}_k+\text{SC}_k)~/~\text{VaR}(\text{DI}_k)$ we find a median of 1.44. The interquartile range of VaR amplification factors ranges from 1.24 to 1.59 with extreme values at 1.11 and 1.87. 

The IB contagion effect on VaR is relatively small for most banks. IB contagion increases the median VaR by 0.42\% of a banks' equity, i.e., the median of VaR(DI$_k$ + IB$_k^\text{WO}$)-VaR(DI$_k$) is 0.42\% (height of the purple hatched bar). The median VaR(DI$_k$+SC$_k$+IB$_k$)-VaR(DI$_k$+SC$_k$) is 0.86\% when supply chain contagion is present (purple bar). 
The amplification factor for the interbank losses, $\frac{\text{IB}^\text{W}_k}{\text{IB}^\text{WO}_k}$ (comparing the solid purple bars with the hatched purple bars), shows that the interbank losses are about doubled. The median value of the amplification factors is 1.91 with a first quartile of 1.70 and a third quartile of 2.08, the extreme values being 1.22 and 2.34. The presence of supply chain contagion increases the size of potential equity losses due to interbank contagion by 91\%.
Interestingly, two of the 19 banks are especially vulnerable to interbank contagion. Bank 7 has no direct exposure to supply network risk and sees its VaR rise from 3.1\% to 5.4\% of its total equity, when supply chain contagion is taken into account. This is a large relative increase of 72\%. Bank 8 shows a particularly high interbank exposure, with the interbank network increasing VaR by 14.9\% ($\text{VaR}(\text{DI}_8+\text{SC}_8+\text{IB}_8^\text{W})-\text{VaR}(\text{DI}_8+\text{SC}_8) = 14.9\%$). Without supply chain contagion, interbank connections alone increase losses by 7.6\% ($\text{DI}_8+\text{SC}_8+\text{IB}^\text{WO}_8-\text{DI}_8 = 7.6\%$). IB contagion is relatively important for bank 8 and supply chain contagion increases the IB contagion effects by 96.1\%.

To see the impact of individual banks' equity losses on the banking system we show the VaR of individual banks weighted by the total equity in the banking system in Fig.~\ref{fig:Fig3_System_SI}. Although banks might only see small or medium individual risks, their size can translate a small individual VaR of only 3.4\% of a banks' equity to a VaR of 1.4\% of equity in the entire banking system. See sec.~\ref{sec:App_SystemicRisk} for further details. 

\subsection{How supply chain contagion amplifies interbank contagion}\label{sec:Fig4}

\begin{figure}[p]
    \centering
    \includegraphics[width=\linewidth]{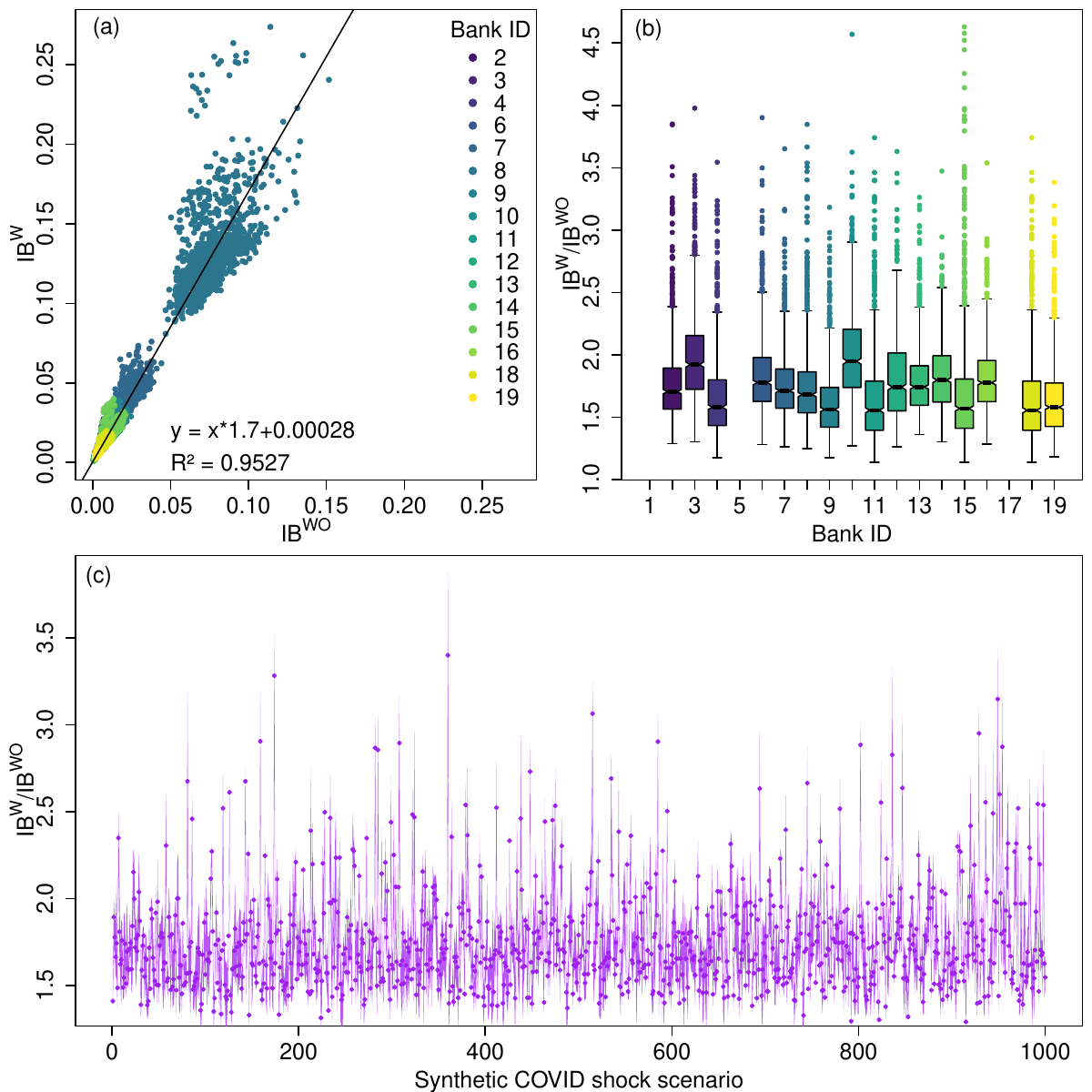}
    \caption{\textbf{Effect of supply chain contagion on losses from interbank contagion.} Panel (a) shows, for 1,000 synthetic scenarios, for each bank, the losses from the interbank contagion without supply chain contagion $\text{IB}^\text{WO}$ on the x-axis and the losses in the same scenario with supply chain contagion, $\text{IB}^\text{W}$ on the y-axis. Every dot corresponds to losses for a single color-coded bank in a single scenario. A linear fit through the data yields a slope of 1.7, meaning that on average the losses on the interbank market are 70\% higher if the supply chain contagion is taken into account. (b) shows box plots of the amplification factor, $\frac{\text{IB}_\text{W}}{\text{IB}_\text{WO}}$, for every bank over the 1,000 scenarios. Most values are close to 1.7 but they can be substantially higher for every bank and reach values of more than 4 in the most extreme cases. Panel (c) shows, for all scenarios the median value of $\frac{\text{IB}_\text{W}}{\text{IB}_\text{WO}}$ as a dark dot, and the range between the first and third quartile as a shaded area. While the values are centered around a value of 1.7 there is a considerable number of scenarios with an amplification of up to 3.5.}
    \label{fig:4}
\end{figure}

Next we examine the impact of supply chain contagion on interbank risks in more detail. Fig.~\ref{fig:4}~(a) shows the marginal interbank equity losses of all banks with supply chain contagion ($\text{IB}_k^\text{W}\left(\psi\right)$) against the marginal interbank equity losses without it ($\text{IB}_k^\text{WO}$) for each of the 1,000 COVID scenarios. Each color represents a different bank. Note that, by construction, $\text{IB}_k^\text{W} > \text{IB}_k^\text{WO}$ for all scenarios and banks. Notably, bank 8 stands out as particularly exposed to interbank market risks, showing substantially higher losses compared to others, see upper part of the figure.
The data in Fig.~\ref{fig:4}~(a) seem to follow a linear relation. Using an OLS regression of $\text{IB}^\text{W}$ on $\text{IB}^\text{WO}$, reveals that, on average across all banks, losses resulting from the interbank network increase by a factor of $1.7$, when supply chain contagion is taken into account, with an $R^2$ of $0.95$. That means that losses from the interbank contagion with supply chain contagion increase by 70\% when losses from interbank contagion without supply chain contagion increase by 100\%. A fit to a power law yields an exponent of $0.967$ -- close to $1$, as can be seen in Fig.~\ref{fig:SI4_log}.

For the amplification of the interbank equity losses for each bank across all 1,000 scenarios we calculate the fraction $\frac{\text{IB}_k^\text{W}}{\text{IB}_k^\text{WO}}$. In Fig.~\ref{fig:4}~(b) we show box plots of the distribution of amplification factors. Remarkably, risks from interbank contagion increase by a factor of more than 3, for almost all banks in at least some scenarios. This means that for every bank there are scenarios where supply chain contagion causes a large amplification of losses in the IB contagion channel, and scenarios where supply chain contagion leads only to minor amplification. High IB amplification for a bank occurs in scenarios where SC contagion causes high losses for one of their important IB loan debtors that do not occur from the direct impact alone. These losses then propagate through the interbank connection. We examine the amplification factors of the two largest banks in more detail in Fig.\ref{fig:Bank_Amplification}, where we show the amplification factors of interbank equity losses for each scenario. Even for these important banks, amplifications of more than 2.5 can be observed in more than 1\% of the scenarios. Figure~\ref{fig:Figure4_Amps} shows the complementary cumulative distribution of the interbank amplifications for all banks over all scenarios. As is common in networked systems, the distribution decays slowly. This shows that the observed large amplifications are not statistical anomalies but extreme events that arise through the interaction of two networked systems. This interaction leads to unexpectedly high systemic losses. Note that the initial shocks all have the same aggregate size, the network structure interacting with the contagion mechanisms and the two contagion mechanisms amplifying each other may yield extreme losses in a few cases. 

Figure~\ref{fig:4}~(c) shows the amplification factor for IB induced bank equity losses for each of the 1,000 scenarios. We show the median value of risk amplification factors over all banks for each scenario as a dark dot and the interquartile range as a shaded area. We disregard banks without interbank losses, since their interbank amplification $\frac{\text{IB}_k^\text{W}}{\text{IB}_k^\text{WO}}$ is not defined. A majority of the interbank risks are increased by a factor close to 1.7 that we calculated from panel (a). However, a few scenarios contain banks whose interbank losses increase by a factor of 3 and more once supply chain contagion is taken into account. This highlights the heterogeneous effect that supply chain contagion can have on interbank contagion and the resulting equity losses.

Note that the apparent linear relation in Fig.~\ref{fig:4}~(a) occurs in the system wide regression due to the particular interbank network used and due to the relative size of $\text{DI}_k$, compared to $\text{DI}_k+\text{SC}_k$. We conduct a series of sensitivity analyses in Sec.~\ref{sec:App_Robustness}. In Sec.~\ref{sec:Model_assumptions} we explain in detail how the linear relation occurs as a result of the interbank network and the shock magnitudes. Note that, when regressing on individual banks, the linear relation does not hold, see Sec.~\ref{sec:Bank_regression}.

\section{Discussion}

To identify the weakpoints of the stability of the financial system, a rigorous understanding of systemic risk, the various contagion channels causing it, and the interaction between these channels is necessary. Here we propose a comprehensive systemic stress testing model featuring interbank contagion, firm-bank contagion, and firm-firm supply chain contagion, which allows us to study how these channels interact. All contagion channels are calibrated to a unique micro-data set containing supply chain relations, bank-firm loans, and interbank loans. Using firm and bank data from Hungary for the year 2019 we can reconstruct the country's entire supply chain network with more than 240,000 firms, their inter-firm connections, and financial statements for most of these entities. With the financial information of firms we can model their default dynamics as a response to specific shocks in a simple, yet highly realistic, manner. If a shock is significant enough to reduce a firm's equity or liquidity to zero it defaults on any liabilities towards financial institutions. Using data from the stress testing framework of the Central Bank of Hungary, we calculate the resulting bank equity losses for the 19 relevant banks. Finally, we consider further indirect losses through an interbank solvency contagion mechanism.

Within this framework we addressed three topics. First, we assessed the systemic financial risk posed by individual firms to the financial system, through direct losses and indirect losses from supply chain contagion, and examined how interbank contagion amplifies that risk. Our results indicate that, although relatively small in comparison, the interbank network is able to inflate the financial systemic risk of the most critical firms by 12\%. For these highly systemically relevant firms the financial system is exposed to losses of 22\%, increased to 24\% with additional interbank contagion. The magnitude of this increase can be explained by the relative size of the interbank market compared to the total exposure of banks to the production network. The interbank network is only 12.5\% the size of the loan exposure of banks to firms. We showed that losses from systemically important firms are comparable in size to the complete default of the second largest bank in the network, highlighting the potential impact of them. From previous studies \citep{Tab24} we know that these losses are primarily driven by the supply chain contagion mechanism. For firms with a lower systemic impact the interbank network can increase the financial systemic risk by up to 28\%.

Second, we analyzed the sources of systemic risk for the financial system and individual banks using realistically calibrated firm-level shocks that match , as a relevant example, the economy-wide impact of the COVID-19 pandemic. We showed that the value at risk for financial system wide equity losses increases by 58.5\% due to supply chain contagion and by an additional 9.6\% if interbank contagion is considered. The network nature of the SC and IB contributions leads to a fat-tailed distribution of losses, meaning that large losses become much more likely. Further, we showed that the median of the total VaR, including supply chain and interbank effects, is 48\% larger compared to the simulation without supply chain cascades. Except for banks that are not connected to the supply chain network, the risk from direct firm failures due to the COVID-19 shock is consistently found to be the largest source of risk. Supply chain contagion increases VaR by 44\%, whereas the increase due to interbank contagion is only 7.7\% of the direct losses. The results are similar for EL and ES.

Third, we examined systemic risks that originate from credit exposures in the interbank network and quantified their increase if the underlying shocks to banks are amplified by supply chain contagion. We show that this amplification can be as much as an additional 70\%. From the perspective of individual banks, the risk amplification can reach values of 300\% in 1\% of all observed scenarios. This highlights the extent of hidden risk in the system that are ignored when supply chain networks are not considered as a risk propagation channel. 

In summary, these numbers demonstrate the need for a more comprehensive approach when analyzing financial systemic risk. In particular it is necessary to link financial risks to default dynamics in the real economy, i.e., the supply chains. The traditional view of financial systemic risk in terms of financial exposures only \citep{Con17,Gla16,Wie23}, must be extended. It is known that different layers of financial exposures can significantly increase systemic risk \citep{Pol15}. Here, we demonstrate the necessity of combining financial layers with the layer of the real economy in the form of buyer-supplier networks if one wants to realistically estimate levels of financial systemic risk. Systemic risk has been addressed independently in both, the literature on shock propagation in supply chain networks \citep{Hal08,Pic22,Bor20a,Car21,Ino19,Die22,Bar16a,Tab24,Hur24,Cor19,Die24a,Dem24}, and in the financial contagion literature \citep{All00,Eis01,Bos04,Bos04a,Bat12,Thu13,Pol15,Bat16a,Gla16,Con17,Bar21,Car24,Wie23,Lev15,Fei17,Ach14,Bor14,Det21,Cac15,Sil18,Pol18,Gut20,Bor20a,Ace15}, however, a systematic combination of both has not been possible, due to data and computational issues. We demonstrated that these limitations do no longer necessarily obstruct to study financial and economic contagion channels and their interactions jointly within one framework. It is fascinating that it becomes immediately apparent that indirect losses from supply chain contagion can outweigh losses from interbank connections. The framework allows one to find the reasons behind these results. Here the high contributions from the SC channel is subject to the relative size of banks' exposures to these channels; the average bank has a ten times larger exposure to firm credits than to interbank loans. 
As a takeaway, in addition to different types of financial interdependencies, such as common asset holdings and creditor-debtor relationships, we find that supplier-buyer connections of banks' clients in the real economy are a most considerable source of financial systemic risk that is capable of inflating the risk transmitted through the interbank layer by 70\%. We have also shown that the interbank layer can increase the financial systemic risk single firms pose to the financial system by up to 28\%. 

Our framework could be seen as a novel tool for comprehensive risk management and stress testing that explicitly takes the temporal details of interconnected systems into account. In that sense the framework is fully data-driven. 
For comprehensive financial stress testing the method of generating direct shocks to firms of the same aggregate size \citep{Die24} with subsequent supply chain contagion can be used as a framework to generate supply chain adjusted shock scenarios for the financial system. Policy makers can choose how strong different sectors would be affected from specific exogenous to the real economy shocks (e.g., trade war, war, natural disaster, pandemic). the presented method generates synthetic firm-level shocks corresponding to the aggregated shocks that were used as an input. In that way our method produces a comprehensive and self-contained financial stress test framework including supply chain contagion.

\paragraph{Limitations}
We list a few limitations to this study. The supply chain shock propagation model employed here follows \citep{Die22} and is designed for short-term shocks and stress-testing purposes. As an input it uses firm-specific production functions that have to be estimated from various data sources. In these estimates, in particular in making decisions on whether an input of a company is essential for production or not, we expect to make errors. We know that wrong assignments can have strong effects, but we have --at the moment-- no possibility to estimate the size of these errors. A second input to the shock propagation model is the relative 'ease' with which firms can replace suppliers. Currently we estimate this 'replaceability' by the inverse of market shares of firms outputs. It is needless to mention that these market shares are hard to estimate since we do not have explicit knowledge on the products that firms produce. Also, we do not take into account geographic distances or historic trading relations into account for estimating the 'replaceability'. Again, it is hard to estimate the relevance of the such introduced errors. Only more granular data and a deeper analysis of the actual replacing dynamics of firms will help to ameliorate this situation. 

A further obvious shortcoming is that due to the lack of complete international firm-level data, our supply chain propagation model can not cover international transactions. The availability of alternative international sources for specific goods and services would of course reduce contagion effects in the supply chain network as we study it. In that sense we can interpret our results on systemic risk quantities such as FSRI as upper bounds. This data border problem also plays into the problem of estimating realistic numbers for 'replaceability'.

While our definition of a firm default is reasonable, we assume a loss given default of 100\%. That is of course not realistic and it was shown that losses from the supply chain network, either directly or indirectly, are roughly proportional to the implemented LGD \citep{Tab24}. This means that if the LGD of firms say in a sector is 50\% then the losses of banks from the supply chain network are half as large. However, if estimates for LGD for firms or industry sectors are available, this problem can be immediately solved. Again this means that the reported risk values should be seen as upper bounds. The same argument also holds for the linearized DebtRank version that we use. 

In reality, defaults could be prevented by government, firm, or bank interventions, such as it happened in thousands of occasions during the COVID-19 crisis worldwide. This is of course not captured in the presented models. However, the framework can be used to also simulate such interventions. In \citep{Tab24} it was shown that with an additional liquidity support of 0.5\% of total equity in the financial system, the equity losses in the financial system could be reduced by 5\%, a factor of 10. 

A further limitation is that we limited ourselves to a single contagion channel in the financial network, ignoring others that are known to be important \citep{Pol15,Pol21}. The effects of interacting contagion channels between financial institutions has been studied extensively, so here we placed the focus on the relative effects of supply chain contagion on a single, though the most relevant in European economies, financial channel. A possible extension could include a coupling in the bank-to-firm direction by restricting the availability of loans and thus liquidity to firms. While research in this direction exists \citep{Hur24, Dem24, Bor20a, Sil18}, the study of the full two-layer structure of the coupled economic-financial system is still missing. 

\paragraph{Conclusion}

Supply chain dynamics play an unexpectedly important role for financial systemic risk. Supply chain contagion interacts with other financial contagion mechanisms and can amplify interbank contagion. To fully understand the impact of supply contagion on financial systemic risk, it is essential to jointly consider interbank contagion and other financial system contagion channels. The interactions between the real economy and financial systems requires careful attention of the details of the temporal aspects of the underlying networks. Otherwise it is impossible to capture the highly non-linear consequences of these interdependencies accurately. Failing to account for these interactions leads to a drastic underestimation of financial systemic risk.

\section{CRediT author statement}
\textbf{Jan Fialkowski}: Writing – review \& editing, Writing – original draft, Visualization, Validation, Methodology, Investigation, Formal analysis, Data Curation, Software. 
\textbf{Christian Diem}: Writing – review \& editing, Writing – original draft, Validation, Supervision, Project administration, Methodology, Conceptualization. 
\textbf{András Borsos}: Resources, Investigation, Data curation. \textbf{Stefan Thurner}: Writing – review \& editing, Writing – original draft, Supervision, Project administration, Conceptualization.
\section{Funding sources}
This research was funded in whole or in part by the Austrian Science Fund (FWF) [grant no. 10.55776/I5985] and the Jubilaeumsfonds of the Austrian central bank project under P18696 and UK Research and Innovation MSCA Postdoctoral Fellow guarantee under EP/Z003199/1.
\section{Data availability}
The data are only physically available in the Central Bank of Hungary or the Hungarian Academy of Sciences and, hence, cannot be shared by the authors. The code to reproduce this study is available on \url{https://github.com/JanFialkowski/FSRI_Plus}.
\newpage
\bibliographystyle{elsarticle-harv}\biboptions{authoryear}
\bibliography{FSRI_Plus}
\clearpage


\appendix

\section{Financial Systemic Risk}
\subsection{DebtRank of banks}\label{sec:App_DebtRAnk}
\begin{figure}
    \centering
    \includegraphics[width=\linewidth]{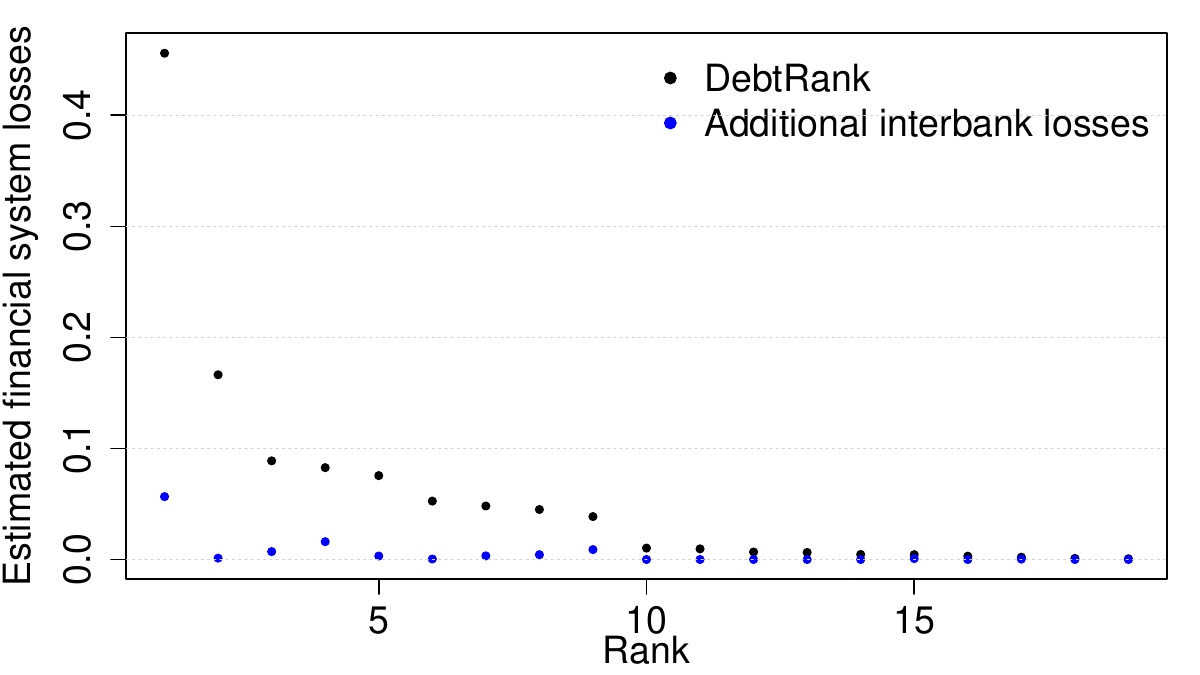}
    \caption{This figure shows the DebtRank according to \citep{Bat12} for each bank with black dots. The blue dots show only the additional losses in the financial system due to contagion effects, without the equity of the originally failing bank.}
    \label{fig:Fig2_DebtRanks}
\end{figure}
Fig.~\ref{fig:Fig2_DebtRanks} presents the DebtRank for each of the 19 banks, as defined in \citep{Bat12}. This measure represents the total loss of equity in the financial system if a single bank were to completely default. The black dots show the total equity lost relative to the financial system's overall equity, while the blue dots indicate additional losses specifically due to contagion in the interbank network — i.e. losses excluding the equity of the initially failing bank. The most impactful bank’s default would result in approximately 45\% of equity losses, with 40\% being its own equity and only 5\% arising from interbank contagion.

This reveals two key aspects of the interbank network. First, although this group represents the largest banks in Hungary, the majority of the financial system's equity is concentrated in just two banks. Second, the total size of the interbank market accounts for only a small portion of the financial system's equity. Consequently, contagion effects within the interbank layer are small compared to losses stemming from the production network.

\subsection{Analysis of FSRI amplification}\label{sec:App_FSRIAmp}
\begin{figure}
    \centering
    \includegraphics[width=\linewidth]{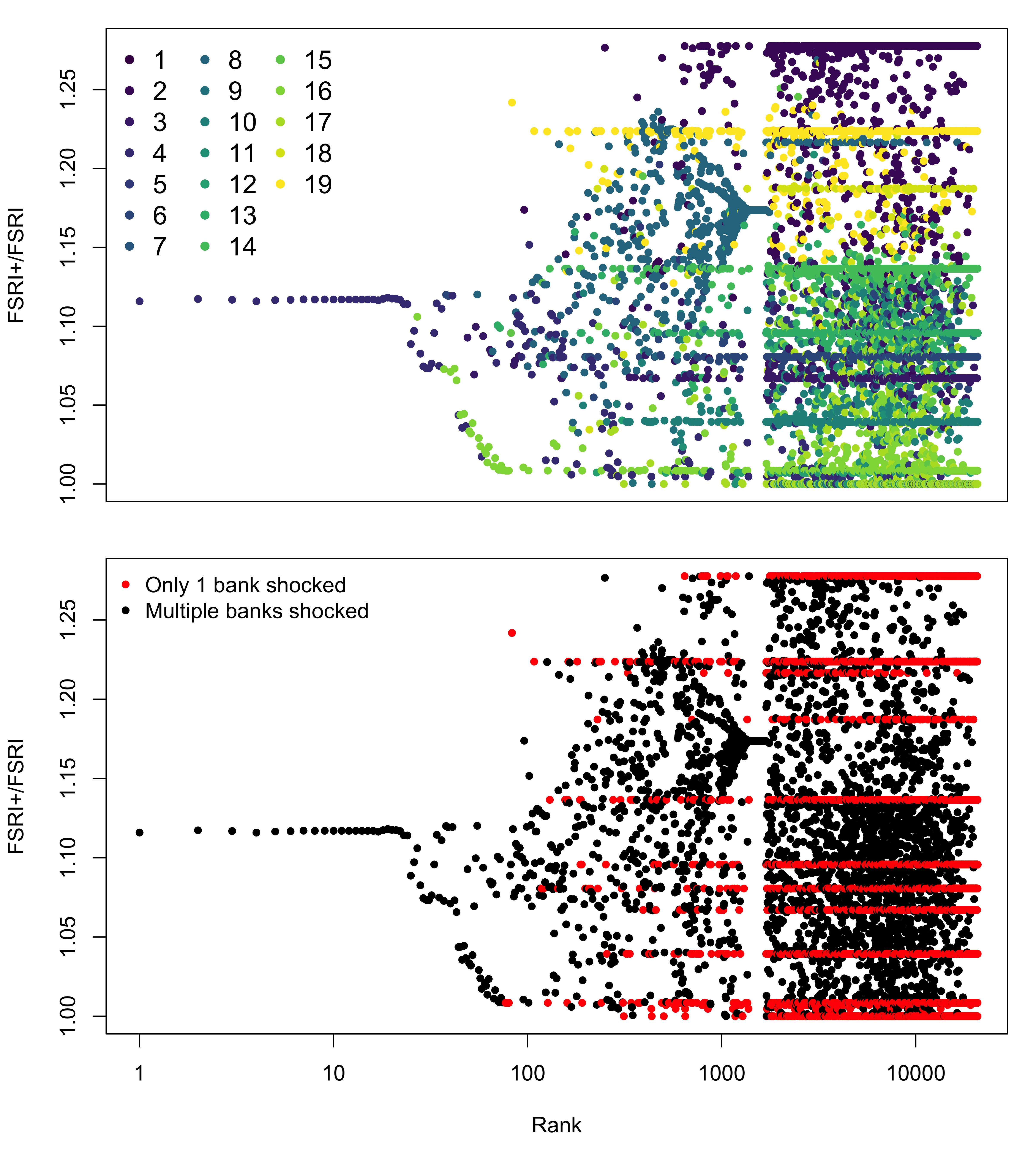}
    \caption{Amplification factors $\frac{\text{FSRI}^+_j}{\text{FSRI}_j}$ for each individual firm. Top: color denotes which bank experiences the highest initial losses before the interbank contagion, bottom: Red dots denote scenarios where only a single bank incurs losses after the supply chain propagation. It can easily be seen that the observed plateaus of the amplification factor arise from single banks being shocked, the linear nature of the employed interbank contagion algorithm with the small size of the resulting shock of banks' equity gives rise to fixed amplification factors.}
    \label{fig:FSRI+colored}
\end{figure}
The horizontal stripes in the distribution of amplification factors for firms' Financial Systemic Risk is caused by these lower ranked firms causing losses for only a single specific banks. Fig.~\ref{fig:FSRI+colored} shows this clearly. The upper graph color codes the bank that is affected the most by the failure of a particular firm, including the supply chain contagion. It can be clearly seen, that each horizontal stripe has a different color, meaning that for every stripe a different bank is most affected. Further, the bottom part shows with a red dot, the firms that cause losses in exactly one single bank. The horizontal stripes are clearly all scenarios in which only one bank is affected. Together both graphs clearly show the cause for the horizontal stripes. 

If a single bank is shocked and the shock is small enough to cause only a small cascade on the interbank market, it is possible to calculate the exact Amplification factor. Let the initial shock to banks be denoted $\psi$ with $\psi_i=\psi$ for $i=k$ and $\psi_i=0$ otherwise, i.e. only bank $k$ receives an initial shock. For a small shock, with only a single propagation step on the interbank network, the resulting Financial Systemic Risk Indices can be calculated as follows
\begin{align*}
    \text{FSRI} &= \frac{e_k\psi}{\sum_i e_i} \\
    \text{FSRI}^+ &= \frac{e_k\psi_k + \sum_j e_j\Lambda_{kj}\psi}{\sum_i e_i}.
\end{align*}
With the definition $\Lambda_{kj}=\frac{L_{kj}}{e_j}$ we can derive
\begin{equation*}
    \frac{\text{FSRI}^+}{\text{FSRI}} = 1+\frac{\sum_j L_{jk}}{e_k}.
\end{equation*}
This clearly shows that if the failure of a firm with the subsequent supply chain contagion affects only a single bank, then the resulting amplification factor depends on $\frac{\sum_j L_{jk}}{e_k}$.

\subsection{Distribution of Financial Systemic Risk}\label{sec:App_FSRIDistri}
\begin{figure}
    \centering
    \includegraphics[width=1\linewidth]{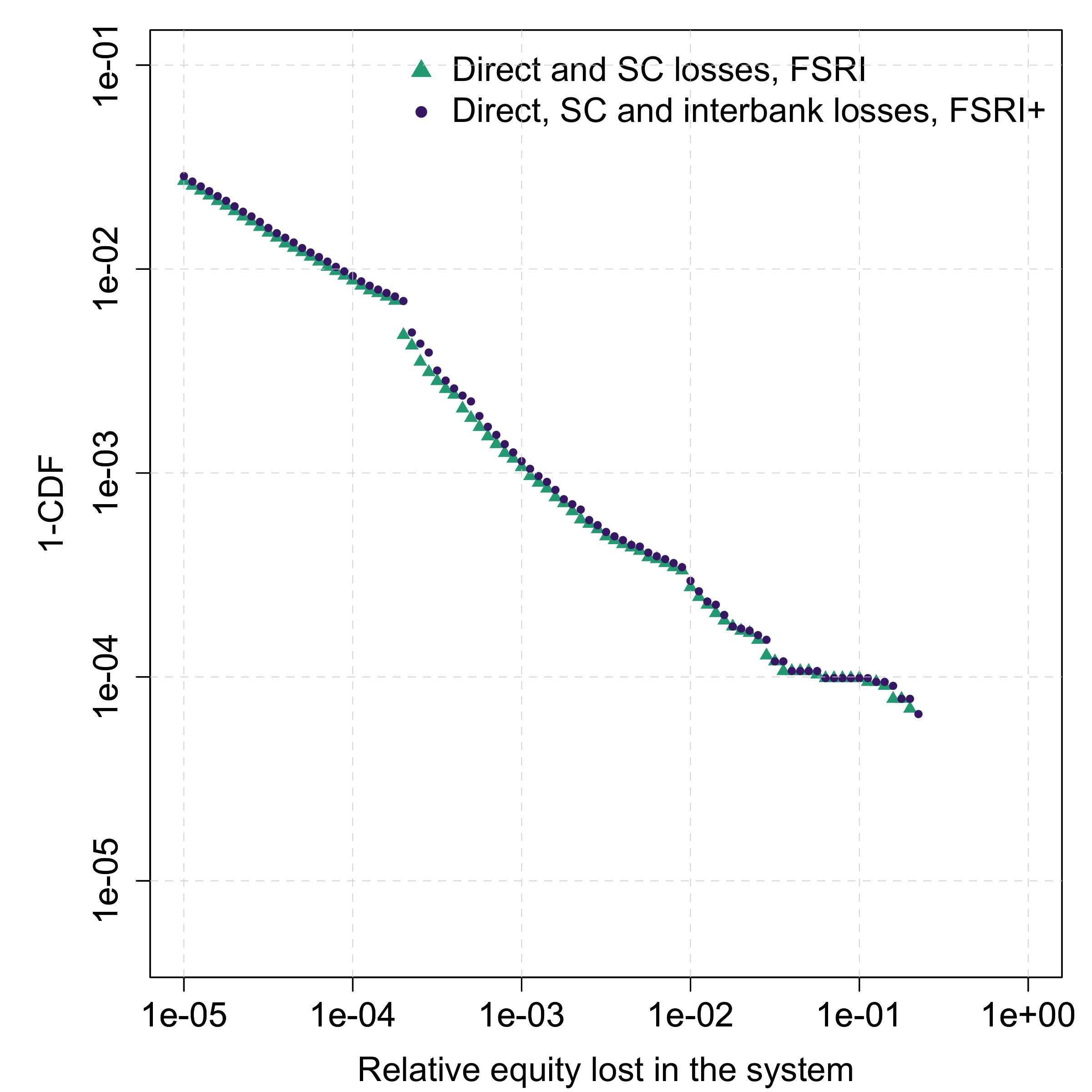}
    \caption{The complementary cumulative distribution function of the financial system losses for single firm failure scenarios. FSRI, without Interbank contagion (teal) and FSRI+ (purple) with the Interbank contagion.}
    \label{fig:Fig2_Survival}
\end{figure}

While the distribution of the Financial Systemic Risk Index does not follow a pure power-law it is nevertheless heavy tailed. Fig.~\ref{fig:Fig2_Survival} shows the complementary cumulative distribution function of FSRI and FSRI$^+$. The addition of the interbank contagion channel for FSRI$^+$ slightly shifts its distribution to the right, meaning it is more likely to see higher losses. The plateau is clearly visible for both curves above losses of 10\% of the financial systems equity. But even for firms not in the plateau surprisingly high financial impacts are possible, 0.1\% of firms are capable of impacting 0.1\% of the financial systems equity. That is a single firm ceasing production and causing losses in the financial system through supply chain and interbank contagion. 

\clearpage
\section{Bankwise Risks under COVID-like scenarios}\label{sec:App_bankwiserisk}

Here we present the results for the equity losses of banks for the risk measures Expected Loss (EL) and Expected Shortfall (ES). We also examine the effect that individual banks' equity losses have on the financial system weighting each banks' VaR with the equity of the financial system.

\subsection{Further Risk measures}\label{sec:App_RiskMeasures}
\begin{figure}
    \centering
    \includegraphics[width=\linewidth]{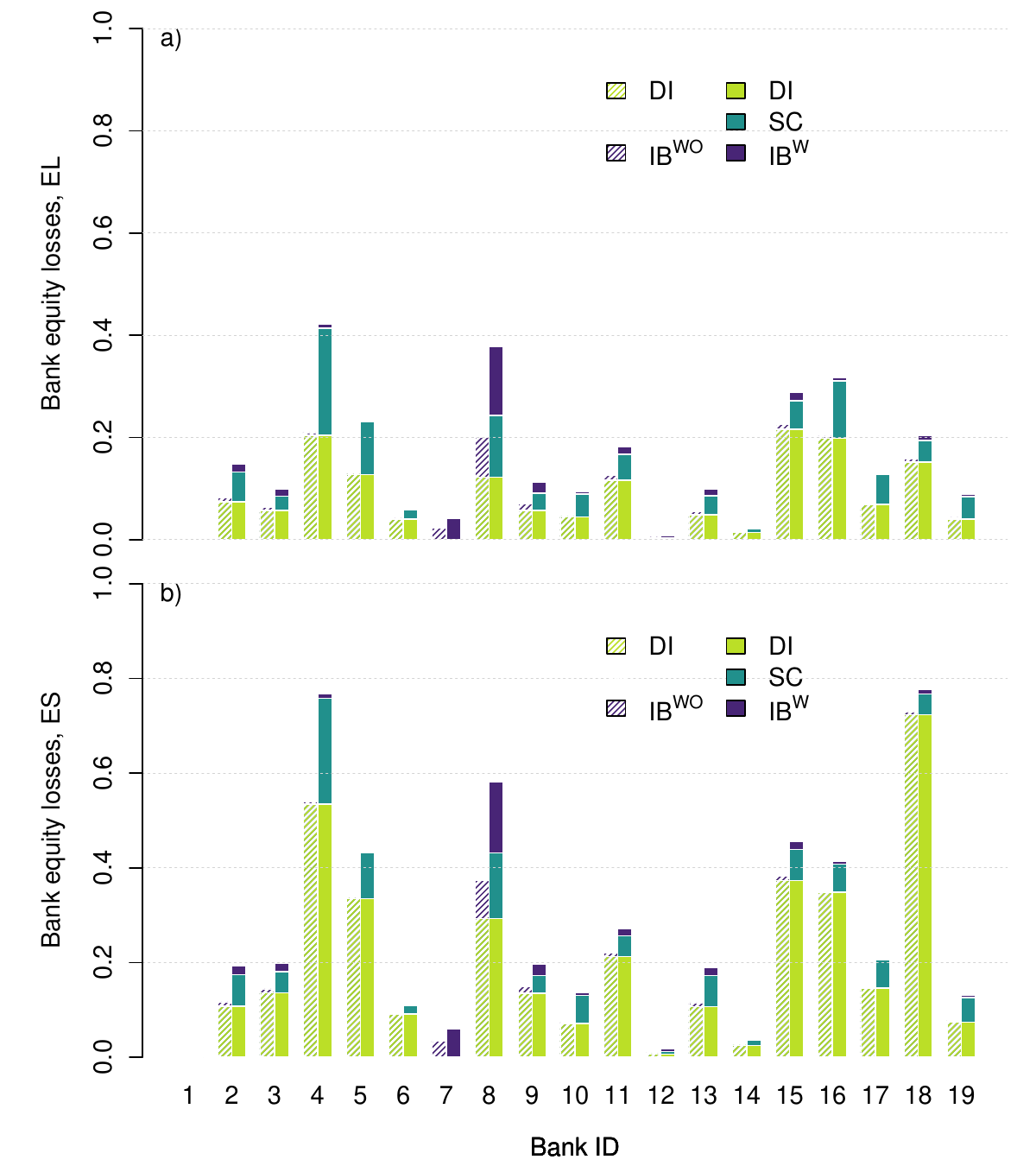}
    \caption{a) The Expected Loss (EL) and b) the Expected Shortfall (ES) for each of the 19 banks. Green: from direct losses, teal: marginal losses from the supply chain contagion and purple: marginal losses from the interbank contagion. The dashed columns correspond to the scenario without additional contagion on the supply chain layer, the solid bars correspond to the scenario incorporating supply chain dynamics.}
    \label{fig:Fig3_EL_ES_SI}
\end{figure}

Besides Value at Risk (VaR), Expected Loss (EL) and Expected Shortfall (ES) are two often used measures for risk profiles. EL is defined as the the mean over all scenarios, and ES is the mean over all scenarios, where the losses are greater than the Value at Risk at a given risk level. We use the 95\% level for Var and ES. For each bank we show the EL in Fig.~\ref{fig:Fig3_EL_ES_SI}~a) for each of the 1,000 COVID scenarios. Without (shaded) and with supply chain contagion (solid) with the color coding for the source of credit risk. From bottom to top, green for direct losses from firm failures, teal represents additional risk due to supply chain contagion and purple denotes additional losses through the interbank network. ES is shown in Fig.~\ref{fig:Fig3_EL_ES_SI}~b). 

The interquartile range of the EL of individual banks' equity losses without supply chain contagion ranges from 4.2\% to 14.2\% with a median value of 6.9\%. With supply chain contagion this increases to 7.5\% and 21.7\% with a median of 11.2\%. The ES of individual banks' equity losses has an intequartile range from 7.5\% to 34.3\% with a median value of 14.3\% without supply chain contagion. Adding supply chain contagion increases the interquartile range of equity losses to range from 12.2\% to 42.3\% with a median value of 19.7\%.

By definition ES is higher than EL, but by how much varies strongly for banks. Without supply chain contagion ES is larger than EL by a factor of 1.9, with factors ranging from 1.4 to 4.6. In the case with supply chain contagion EL and ES are a lot closer, with a factor of 1.6 at the median and ranging from 1.3 to 3.8. Most stylistic facts that could be seen in the VaR are also visible for EL and ES. Namely the disproportionate importance of supply chain shocks for the banks' financial health. The two banks that are sensitive to shocks of the interbank system are also clearly visible.

In sec.~\ref{sec:App_RiskMeasures} we examine Expected Loss (EL) and Expected Shortfall (ES) of banks' equity losses from the 1,000 synthetic COVID scenarios. We observe similar risk patterns, with direct firm defaults being the largest single source of potential risk for banks. Total EL of individual banks' equity has a median value of 6.9\% without and 11.2\% with supply chain contagion, i.e. amplification of 63\%. The median value of the ES of the equity losses is 14.3\% without supply chain contagion and 19.7\% with supply chain contagion, i.e. amplification of 38\%. The interbank network increases Expected Loss by 7\% without and with supply chain contagion, while the Expected Shortfall increases by 3.8\% and 5.1\% respectively, i.e. 34\% amplification.

\clearpage

\subsection{Systemic Risk profile}\label{sec:App_SystemicRisk}

\begin{figure}
    \centering
    \includegraphics[width=\linewidth]{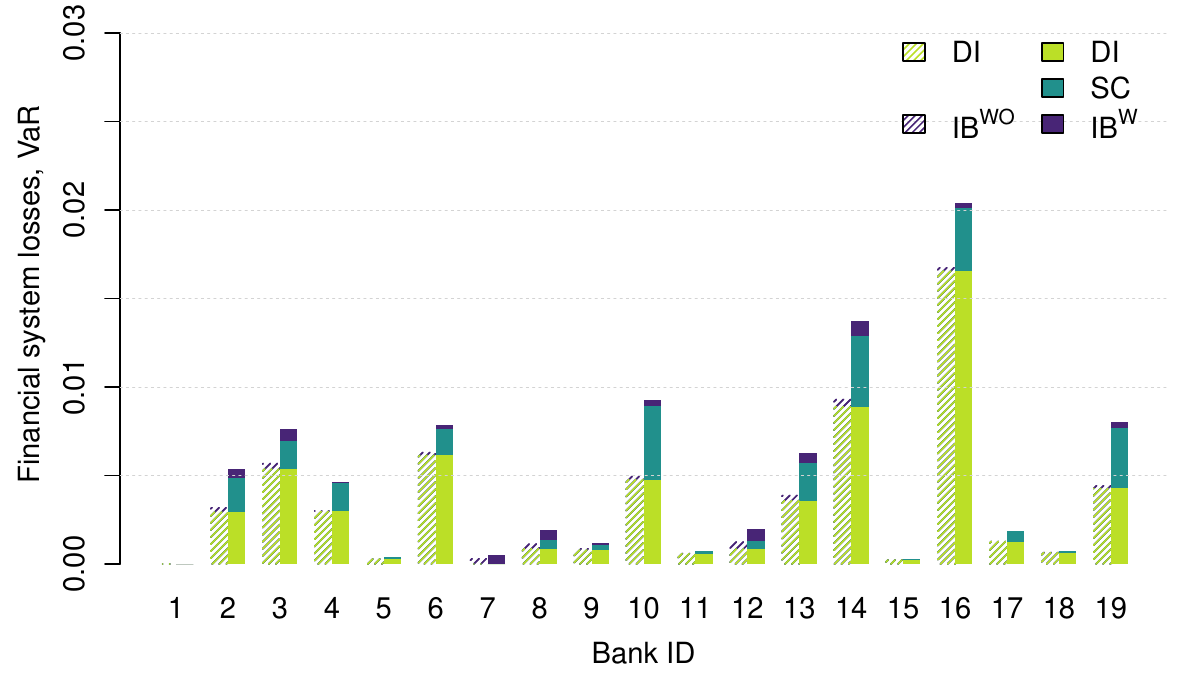}
    \caption{Value at Risk (VaR) in amount relative to the total equity in the banking system. I.e. bank 16 can cause losses of up to 2\% of equity in the entire banking system. Note also that bank 14 causes losses of up to 1.4\% of the financial systems equity, while losing only 3.4\% of its own equity. Total VaR of equity lost in the financial system amounts to 9.3\% with and 6.4\% without interbank contagion.}
    \label{fig:Fig3_System_SI}
\end{figure}

In Fig.~\ref{fig:Fig3_System_SI} we show the Value at Risk for each bank over the 1,000 COVID scenarios. But instead of showing losses relative to a single banks equity, we show losses relative to the equity of the banking system, i.e. Instead of $\text{VaR}\left(\mathcal{L}_j\left(\psi_i,T\right)\right) / e_j$ we plot $\text{VaR}\left(\mathcal{L}_j\left(\psi_i,T\right)\right) / \sum_i e_i$. This allows us to analyze the systemic importance of not only the single scenarios, but also of individual banks. 

Besides the effect of the three different channels for individual banks, losses from direct default, losses due to supply chain contagion and losses due to interbank contagion, we can examine the effect on the equity of the financial system as a whole. When adding the VaR of individual banks weighted by their equity relative to the system-wide equity, our scenarios produce a VaR for the entire banking layer of 6.4\% without and 9.3\% with supply chain contagion, i.e. amplification of 45\%. Notably, a single bank, bank 16, accounts for a system-level VaR of 2\%, one fifth of the total risk. Due to the heterogeneous amounts of equity held by individual banks, a risk that appears small at the individual level for a large bank can contribute significantly to overall systemic risk and vice versa. Bank 14 is such a bank, where a low individual VaR of total equity lost with supply chain contagion of 3.4\% translates into a substantial risk for the entire system of 1.4\%. I.e. this single bank loses only a small amount of its equity but is responsible for 14.8\% of total equity lost in the financial system.

\clearpage
\section{Dependence of interbank risks on Supply Chain Cascades}\label{sec:App_Robustness}

To understand what drives the relatively homogeneously looking amplification across banks Fig.~\ref{fig:4}~a) we conduct a serious of sensitivity analyses.
First, we examine whether the homogeneous amplification can be explained by direct losses plus supply chain losses, $\text{DI}_k+\text{SC}_k$, scaling linearly with the direct losses $\text{DI}_k$. In ~\ref{sec:App_BankShock}, Fig.~\ref{fig:Scenariowise_Bankshock} and Fig.~\ref{fig:Scenariowise_Bankshock_log}, we examine the structure of the shocks that banks receive from the supply chain and show that their relationship is not well described by such a simple linear relationship.
Second, Fig.~\ref{fig:Fig4_Factor} b) shows that the linear relationship of $\text{IB}_k^\text{W}$ and $\text{IB}_k^\text{WO}$ also holds, when correlations between the shocks to banks (induced by the sampling constraints of the synthetic COVID-19 shocks and supply chain contagion) are not present. To show this we use Gaussian distributed shocks with the same mean and variance as the original synthetic COVID-19 shocks. 

Fig.~\ref{fig:Fig4_Factor}~c) shows, that when replacing the empirical interbank network with a randomly generated liability matrix, the linear relationship weakens with $R^2=0.66$. This shows that the particular structure of the interbank network considered here plays an important role in how SC contagion amplifies IB contagion. This is driven by the averaging of the IB network snapshots. Hence, the network structure is important in understanding how financial contagion gives rise to systemic risk, as was shown previously in \citep{Die20}. 
Note that when looking at the risk measures Expected Loss (EL), Value at Risk (VaR) or Expected Shortfall (ES) of the interbank losses $\text{IB}_k^\text{WO/W}$, see Fig.~\ref{fig:Interbank_Amps}, the homogeneous amplification relationship is even more apparent due to the averaging effect these risk measures have. The factors change slightly to 1.73, 1.94 and 1.83 respectively. This is driven through the strong aggregation across scenarios through the risk measures and the homogeneous structure of the IB network.

\subsection{Shocks of the banking system before interbank contagion}\label{sec:App_BankShock}

\begin{figure}
    \centering
    \includegraphics[width=\linewidth]{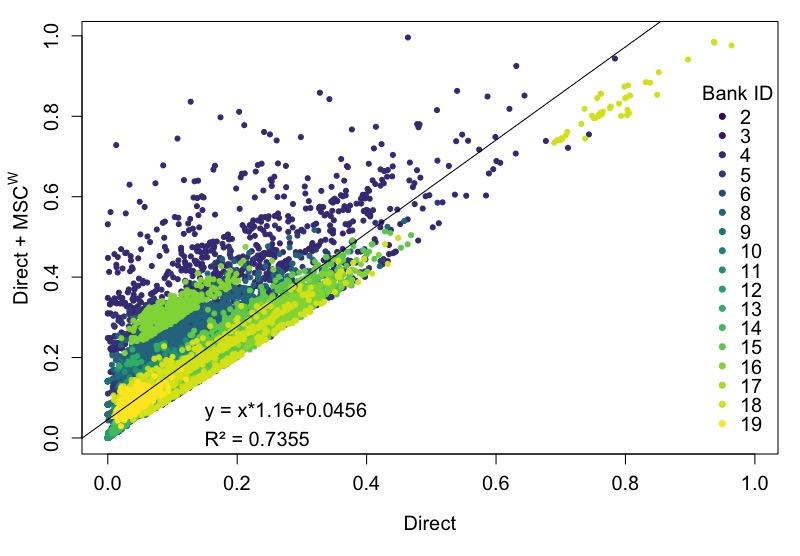}
    \caption{Bank equity losses from firm defaults after the initial synthetic COVID-shocks with supply chain contagion, $\text{Direct}_i+\text{SC}_i$, plotted against the bank equity losses from firm defaults after only the direct COVID-shocks, $\text{Direct}_i$. This is the initial shock for the interbank contagion in the scenario with supply chain contagion against the initial bank shock without supply chain contagion.}
    \label{fig:Scenariowise_Bankshock}
\end{figure}

We examine the shock that banks receive either directly from the original firm-level COVID-type shocks or after a subsequent supply chain contagion. We plot $\text{Direct}_i+\text{SC}_i$ against $\text{Direct}_i$ for each color coded bank $i$ for each of the 1,000 COVID-type scenarios in Fig.~\ref{fig:Scenariowise_Bankshock}. For comparison with Fig.~\ref{fig:4}~a) of the main text we have also added a linear fit through the point cloud. While individual banks tend to be clustered around a line, these shocks alone can not explain the apparent linear relationship between the interbank shocks. From this plot it also easy to identify a small number of scenarios that cause large losses for bank 18, which is also apparent in its Expected Shortfall, see Fig.~\ref{fig:Fig3_EL_ES_SI}~b).

\begin{figure}
    \centering
    \includegraphics[width=\linewidth]{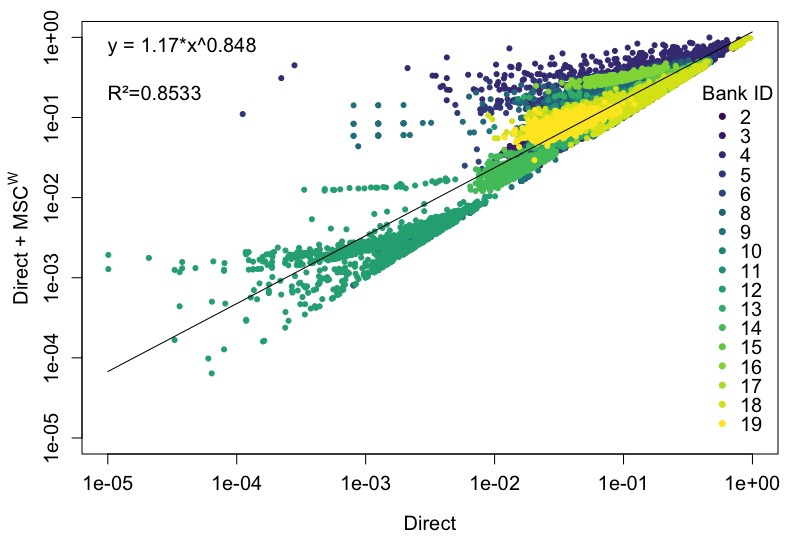}
    \caption{$\text{Direct}_i+\text{SC}_i$ plotted against $\text{Direct}_i$ in logarithmic scale. The horizontal stripes correspond to scenarios, where the equity losses due to direct firm defaults from the synthetic COVID-shocks are small, but supply chain propagation adds additional risk.}
    \label{fig:Scenariowise_Bankshock_log}
\end{figure}

In Fig.~\ref{fig:Scenariowise_Bankshock_log} we show the shock that banks receive in COVID-type scenarios with and without supply chain contagion on a logarithmic scale. While the interbank losses were strongly centered around a line, the shocks to the banks is more heterogeneously spread. Further, the apparent clustering of dots around some horizontal line on the low end of shocks represents a fixed amount of additional risk that is added by the supply chain contagion.

\subsection{Dependence on modeling assumptions and network structure}\label{sec:Model_assumptions}
\begin{figure}
    \centering
    \includegraphics[width=\linewidth]{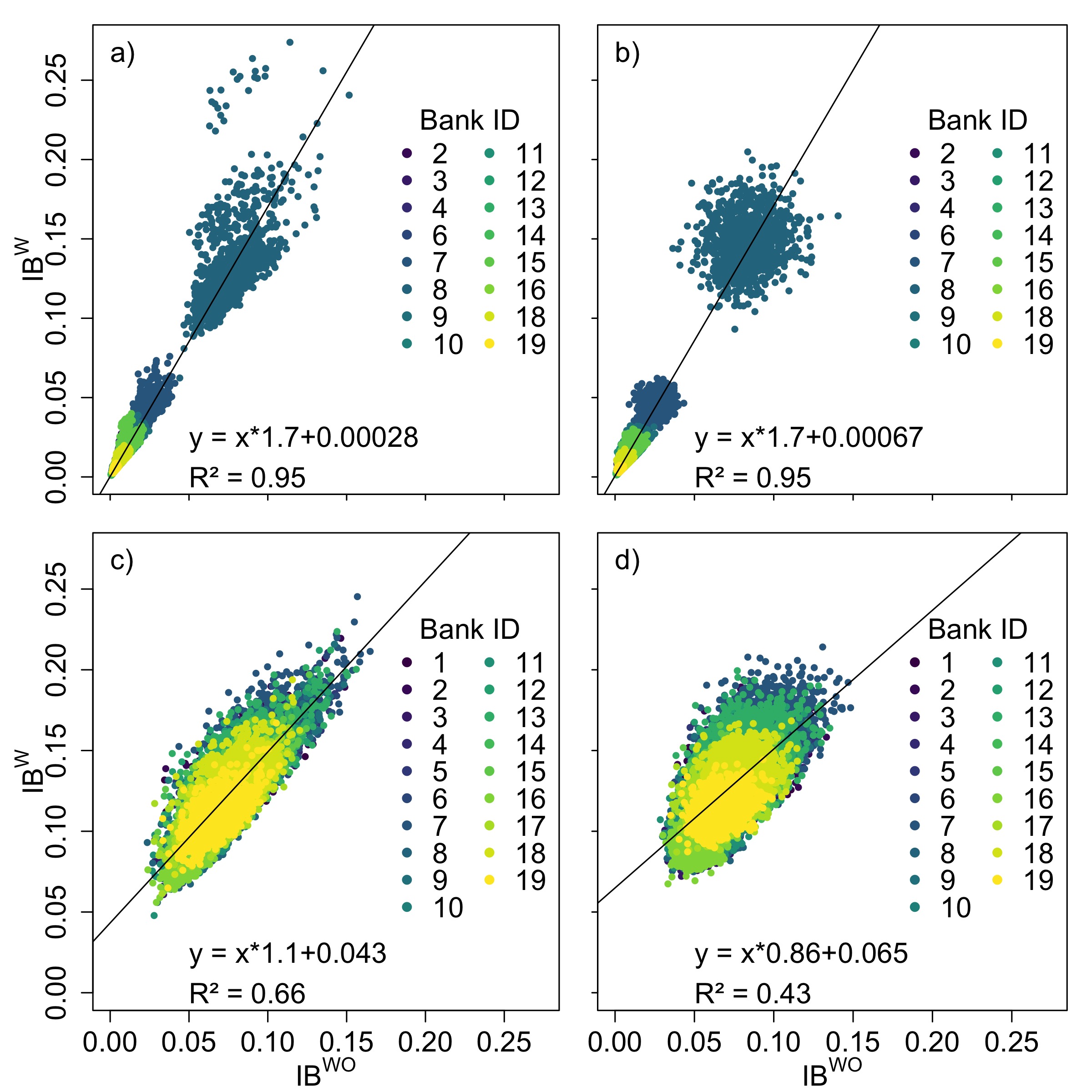}
    \caption{The marginal interbank losses with a supply chain cascade plotted against the marginal interbank losses without a supply chain cascade. The color of a dot denotes the bank it represents. a) shows the data from the synthetic COVID-shocks with the original Interbank Network. It is the same as \ref{fig:4}~a). In b) we draw shocks for the banks from normal distributions with the same mean and variance as for the synthetic COVID shocks. I.e. $\mathcal{L}\left(\Phi_j,t_1\right)=\mathcal{N}\left(\mu_i,\sigma_i\right)$ with $\mu_i$ and $\sigma_i$ calculated from the synthetic shocks. Figure c) shows the original shocks, but with an artificial interbank network. We chose $\Lambda_{ij}=\mathcal{U}\left(0,0.05\right)$ for $i\neq j$ and zero otherwise. This corresponds to each bank borrowing money from each other bank, such that the total exposure to the other banks is at most equal to its equity. Figure d) combines the random shocks with the random interbank exposure network.}
    \label{fig:Fig4_Factor}
\end{figure}

To examine the validity of the results in sec.~\ref{sec:Fig4} we recalculate the interbank contagion part of the model with different shocks from the supply chain and network configurations. Fig.~\ref{fig:Fig4_Factor}~a) shows the same results as in the main text, the additional losses on the interbank market after the 1,000 COVID-type shocks without a supply chain cascade against the additional interbank losses with a supply chain cascade. The most striking feature is that while for each bank the point clouds are fairly widely distributed, in aggregate they can be fit very well by a line described by $y=1.7x$. First we eliminate any correlation between the shocks from the supply chain by drawing 1,000 shocks from a normal distribution with the same mean and variance as the COVID-type shocks. The result of Interbank Contagion with these shocks is shown in Fig.~\ref{fig:Fig4_Factor}~b). The linear fit is unaffected by the change to the shape of the shock. This means the fit is a result of the particular network structure, namely banks' exposure to potential losses on the IB network, and the shocks sizes. The shape of the point clouds for individual banks is now circular, due to the shocks being purely Gaussian. The correlations introduced due to the sampling of the synthetic shocks and SC contagion shapes the shock, the size of the shock and banks' interbank exposures result in the apparent linear relationship. In Fig.~\ref{fig:Fig4_Factor}~c) we use the original COVID-type shocks and instead use a random interbank matrix. We draw random values for each entry $\Lambda_{ij}$ from a uniform distribution $\mathcal{U}\left(0,0.05\right)$ for $i\neq j$. This ensures a maximally random interbank network and destroys network heterogeneities, while ensuring now bank lends to itself or has a total exposure $\sum_j\Lambda_{ij}$ greater than 1. The linear fit has a reduced $R^2$ of $0.66$ instead of the previous $0.95$ with a different slope of $1.1$. This shows that the linear relationship is strongly driven by the network structure, with the shock shape and size enhancing that apparent relationship. Fig.~\ref{fig:Fig4_Factor}~d) shows the results of the interbank contagion with the same random interbank network and the normally distributed random shocks. The linear fit is now even worse with a $R^2$ of only $0.43$. This shows that the shape of the shocks is relevant in shaping the relationship between the interbank losses without and with a supply chain contagion. That shape is itself determined by the underlying production network whose influence is visible even when examining only the additional interbank losses.

\begin{figure}
    \centering
    \includegraphics[width=\linewidth]{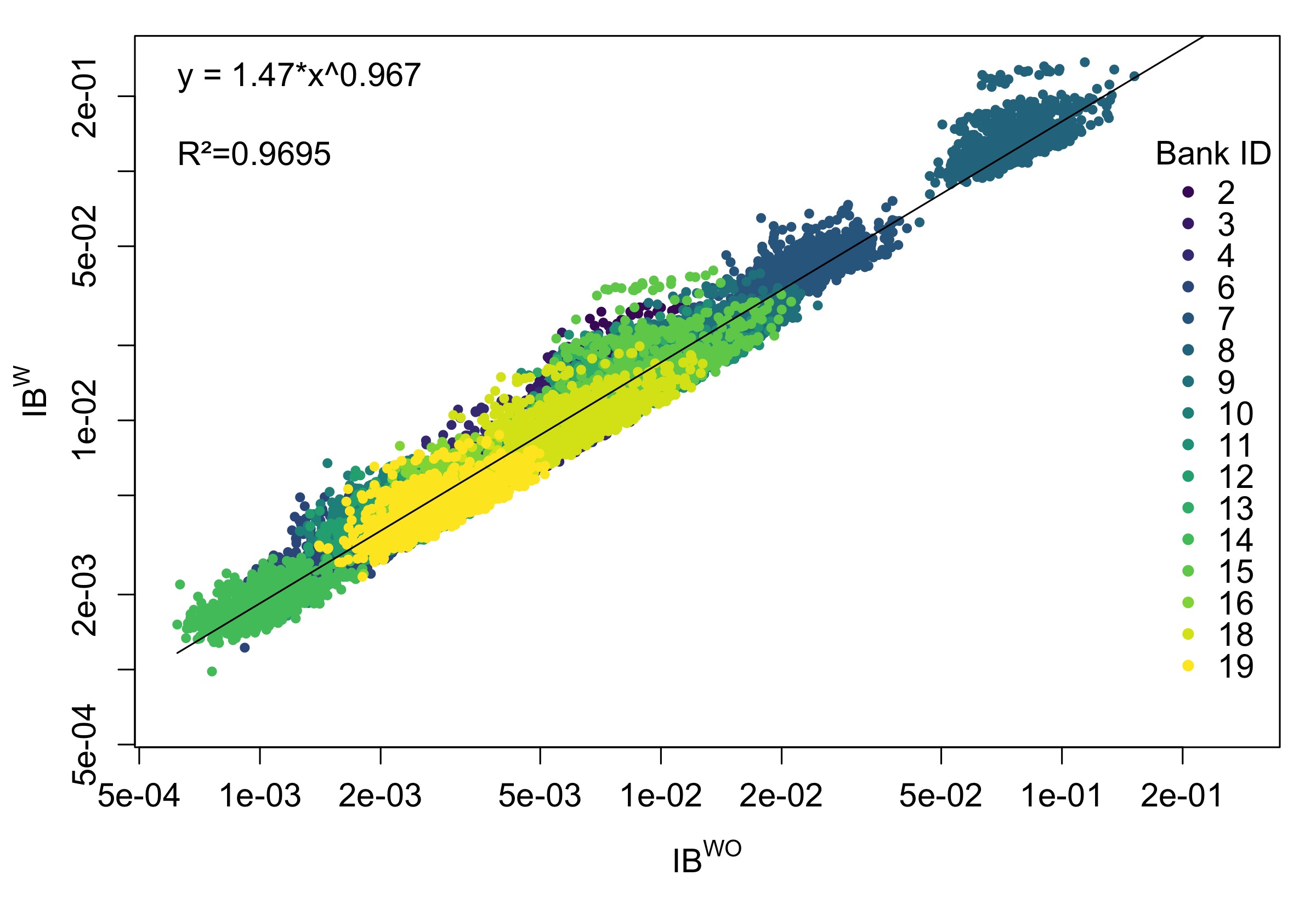}
    \caption{Figure 4~a) with logarithmic axes: For 1,000 simulated scenarios, for each bank, the losses from the interbank contagion without supply chain contagion on the x-Axis and the losses in the same scenario with supply chain contagion on the y-Axis. Each dot corresponds to losses for a single color-coded bank in a single scenario. The line gives the result of a linear fit through the log-transformed variables. The result is a power law with an exponent close to 1, essentially a linear fit.}
    \label{fig:SI4_log}
\end{figure}

The linear relationship between additional interbank losses with and without a supply chain cascade persists even when looking at the log-transformed values. In Fig.~\ref{fig:SI4_log} we plot the results from the main text with log-transformed variables. A linear fit using ordinary least squares regression through the log-transformed data results in a power-law fit. The fitted exponent is close to one with a value of $0.967$ and the slope of is with $1.47$ of the same magnitude as the result from fitting through the non-transformed data. Since the effects or each bank are heterogeneous, see sec.~\ref{sec:App_RiskMeasures} a global fit can only serve as a rough estimate of the increased risk. Thus we report the results of the fit through the non-transformed data.

\subsection{Individual Bank regressions}\label{sec:Bank_regression}
\begin{table}[ht]
\centering
\begin{tabular}{rrrr}
  \hline
id & Slope & Intercept & $\text{R}^2$ \\ 
  \hline
2 & 1.076 & 0.006 & 0.288 \\ 
3 & 1.077 & 0.006 & 0.229 \\ 
4 & 1.010 & 0.003 & 0.463 \\ 
6 & 1.090 & 0.001 & 0.351 \\ 
7 & 1.036 & 0.016 & 0.445 \\ 
8 & 1.046 & 0.053 & 0.293 \\ 
9 & 1.019 & 0.008 & 0.476 \\ 
10 & 0.982 & 0.002 & 0.362 \\ 
11 & 0.998 & 0.005 & 0.498 \\ 
12 & 0.662 & 0.003 & 0.140 \\ 
13 & 1.106 & 0.005 & 0.247 \\ 
14 & 0.938 & 0.001 & 0.406 \\ 
15 & 1.079 & 0.006 & 0.325 \\ 
16 & 0.982 & 0.002 & 0.312 \\ 
18 & 0.998 & 0.004 & 0.498 \\ 
19 & 0.999 & 0.002 & 0.458 \\ 
   \hline
\end{tabular}
\caption{The results of OLS regression of $\text{IB}_k^\text{W}$ against $\text{IB}_k^\text{WO}$ for each individual bank $k$. We present the intercept and slope for each banks regression as well as the value of $\text{R}^2$.}
\label{tab:BankwiseRegression}
\end{table}

Repeating the same Regression for individual banks as was done for the entire system shows clearly that the linear relationship in Fig.~\ref{fig:4}~a) is an aggregate effect, see Table~\ref{tab:BankwiseRegression}. The slope for individual banks is close to one, with a mean of 1.01, and non-negligible intercepts. The average value of $\text{R}^2$ is 0.362. For the individual banks the SC contagion seems to add a fixed amount of risk. Not that three banks are not connected to the interbank network in our sample and are omitted from the regression analysis.

\subsection{Aggregated Measure for Interbank Risk}\label{sec:App_AggregatedRisk}
\begin{figure}
    \centering
    \includegraphics[width=\linewidth]{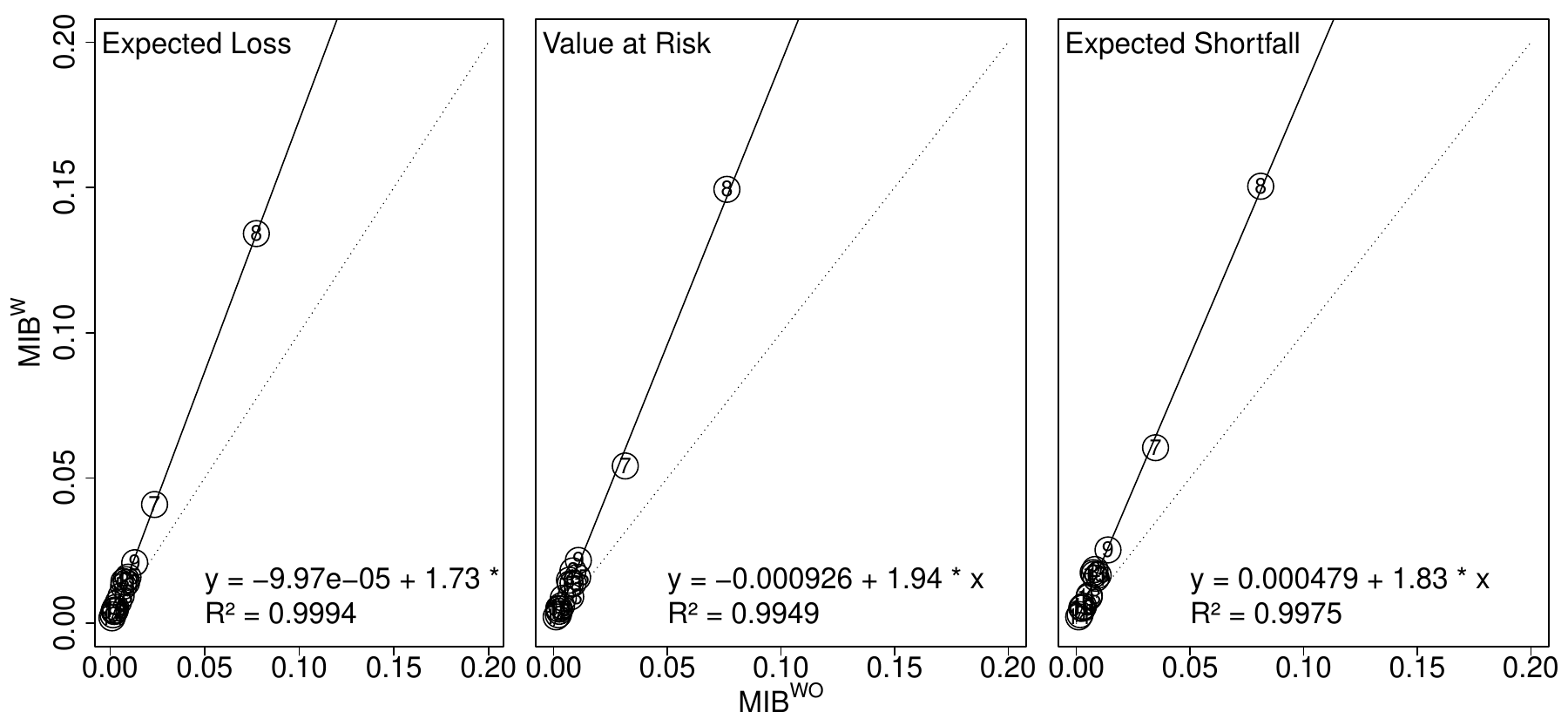}
    \caption{Additional risks incurred from the interbank layer contagion after the supply chain cascade plotted against the additional interbank layer losses without the supply chain cascade. The dashed line is the identity line as a guide to the eye and the solid line corresponds to the best linear fit to the data. The expected losses, value at risk and expected shortfall are higher by a factor of 1.73, 1.94 and 1.83 respectively, if one takes supply chain contagion into account.}
    \label{fig:Interbank_Amps}
\end{figure}

When the additional interbank losses are aggregated to the risk measures Expected Loss (EL), Value at Risk (VaR) and Expected Shortfall (ES) the linear relationship is still visible. The relationship between the EL on the interbank market without a supply chain cascade and with a previous supply chain cascade can be fit fairly well with a line. This leads to an average amplification of risk by a factor of $1.73$. VaR and ES are similarly well described by a linear fit and give an amplification of $1.94$ and $1.83$ respectively.

\subsection{Amplification distributions}\label{sec:AmpDistri}
\begin{figure}
    \centering
    \includegraphics[width=\linewidth]{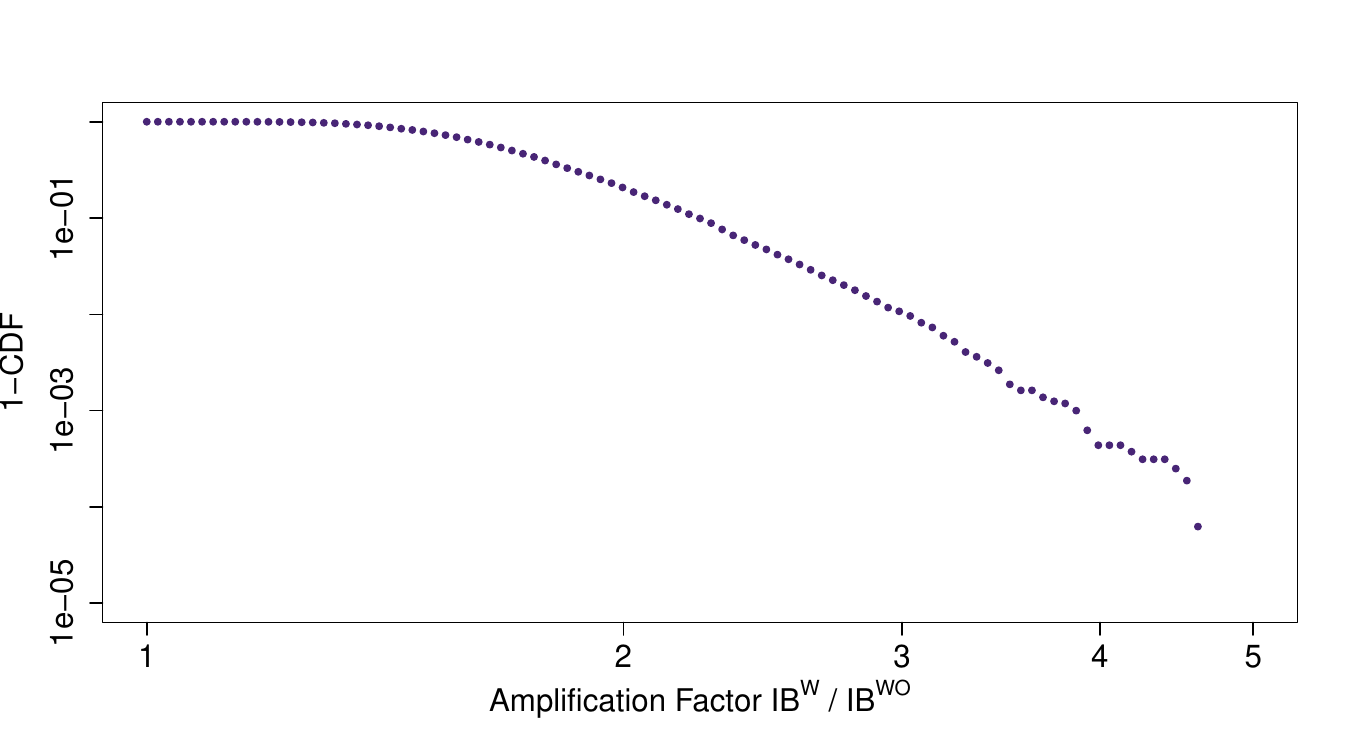}
    \caption{Complementary cumulative distribution function of the Amplification factors $\frac{\text{IB}_\text{W}}{\text{IB}_\text{WO}}$ seen in Fig.~\ref{fig:4}. The distribution is calculated over all banks and scenarios and extends smoothly up to a factor of 4.}
    \label{fig:Figure4_Amps}
\end{figure}

To examine if the observed high amplification factors of additional interbank risks are statistical outliers we plot the complementary cumulative distribution function in Fig.~\ref{fig:Figure4_Amps}. The observed distribution of amplification factors decreases slowly and smoothly from 1 to 4. This means that while the largest amplification factors are rare, they can not be classified as anomalies and need to be taken into account. 

\begin{figure}
    \centering
    \includegraphics[width=\linewidth]{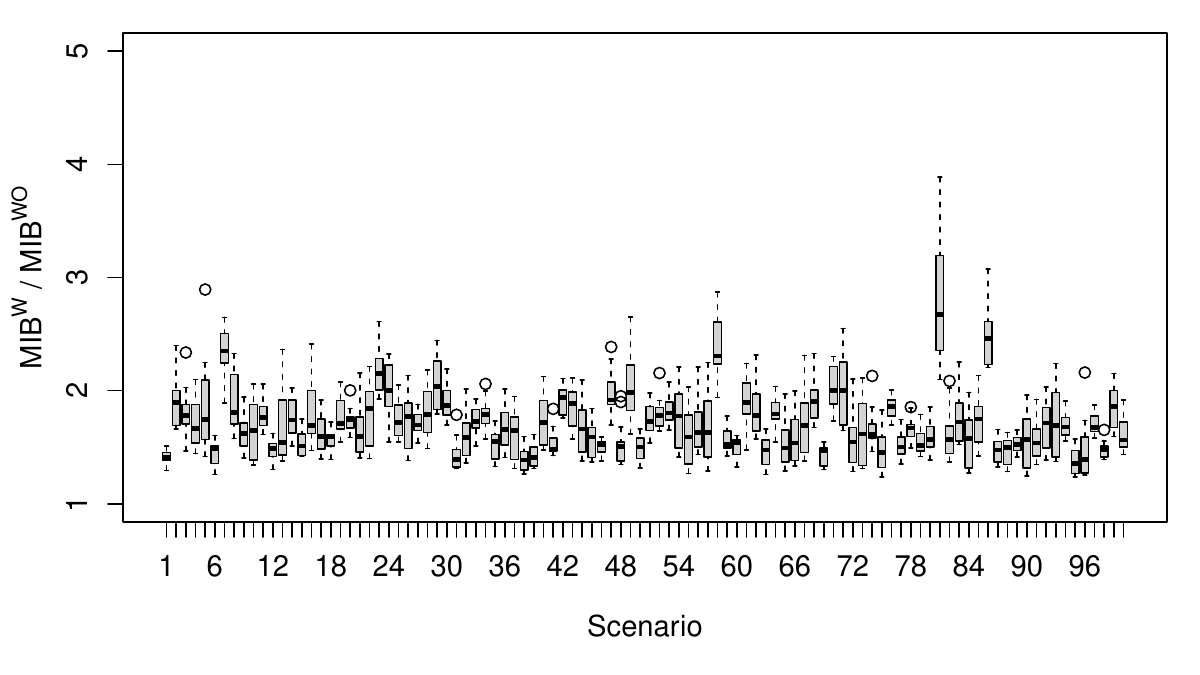}
    \caption{Box plot of relative amplification of marginal Interbank risks for each scenario. For each scenario the banks' amplifications are calculated and the resulting values are represented in the box plot}
    \label{fig:boxplot}
\end{figure}

Instead of showing only the median and interquartile range of interbank amplifications $\frac{\text{IB}^\text{W}}{\text{IB}^\text{WO}}$ for all 1,000 scenarios we present box plots for the first 100 scenarios in Fig.~\ref{fig:boxplot}. The shading in Fig.~\ref{fig:4}~b) of the main text represents only the boxes in this plot. The rare, but high amplification factor of 3 and higher are clearly visible.

\begin{figure}
    \centering
    \includegraphics[width=\linewidth]{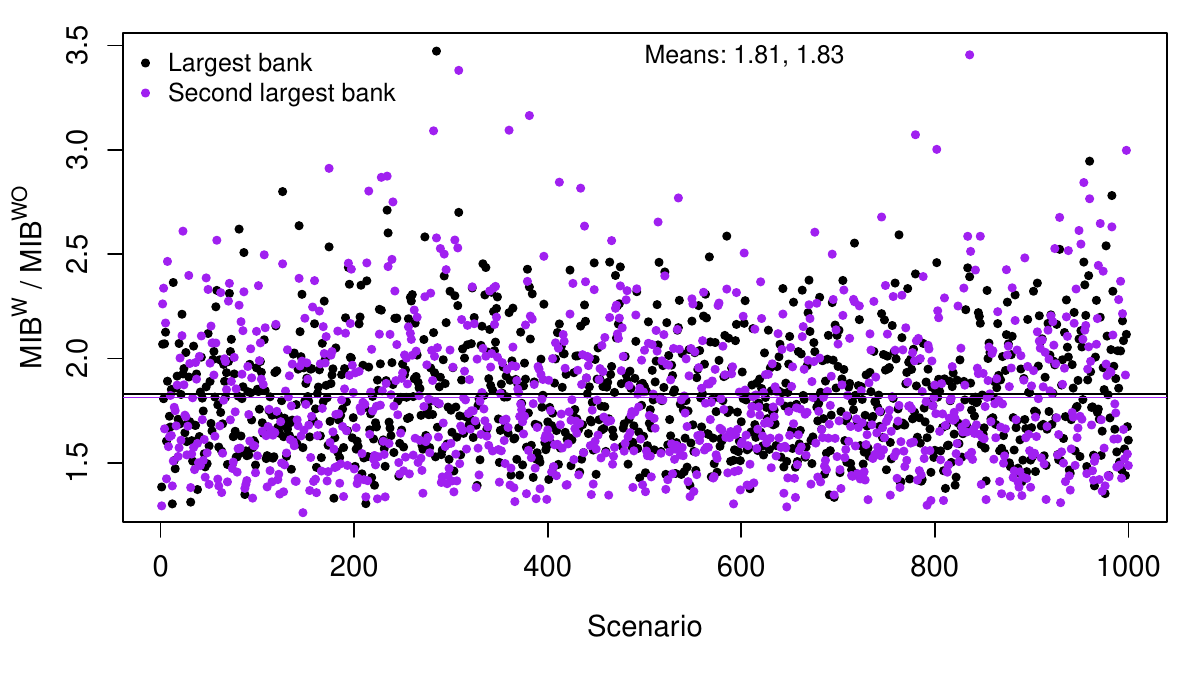}
    \caption{Relative Interbank amplification $\frac{\text{IB}_i^\text{W}}{\text{IB}_i^\text{WO}}$ for each scenario for the two banks with the highest equity in the dataset. Black dots represent the largest and purple dots the second largest bank. Means are marked at 1.81 and 1.83.}
    \label{fig:Bank_Amplification}
\end{figure}

For the two largest banks we calculate the interbank amplification factor $\frac{\text{IB}_i^\text{W}}{\text{IB}_i^\text{WO}}$ for each of the 1,000 scenarios. Together these two banks hold 56\% of the equity in our financial layer. The mean amplification factor for these banks is 1.83 and 1.81 respectively. The distribution is clearly skewed towards high values and scenarios with an amplification of more than $2.5$ can be observed. That means losses through the interbank layer are higher by a factor of $2.5$ than what these banks would predict without taking a possible supply chain contagion into account.

\clearpage







\end{document}